%&plain
%
\let\includefigures=\iffalse
%
% the following is to use blackboard bold fonts --
\let\useblackboard=\iftrue
%
% activate this if you don't have them.
%\let\useblackboard=\iffalse
%
% You might also need to remove this line.
\newfam\black
\input harvmac.tex
\input rotate
\input epsf
\input xyv2
\noblackbox
%\draftmode
\includefigures
\message{If you do not have epsf.tex (to include figures),}
\message{change the option at the top of the tex file.}
\def\figin{\epsfcheck\figin}\def\figins{\epsfcheck\figins}
\def\epsfcheck{\ifx\epsfbox\UnDeFiNeD
\message{(NO epsf.tex, FIGURES WILL BE IGNORED)}
\gdef\figin##1{\vskip2in}\gdef\figins##1{\hskip.5in}% blank space instead
\else\message{(FIGURES WILL BE INCLUDED)}%
\gdef\figin##1{##1}\gdef\figins##1{##1}\fi}
\def\DefWarn#1{}
\def\figinsert{\goodbreak\midinsert}
\def\ifig#1#2#3{\DefWarn#1\xdef#1{fig.~\the\figno}
\writedef{#1\leftbracket fig.\noexpand~\the\figno}%
\figinsert\figin{\centerline{#3}}\medskip\centerline{\vbox{\baselineskip12pt
\advance\hsize by -1truein\noindent\footnotefont{\bf Fig.~\the\figno:} #2}}
\bigskip\endinsert\global\advance\figno by1}
%%%
\else
\def\ifig#1#2#3{\xdef#1{fig.~\the\figno}
\writedef{#1\leftbracket fig.\noexpand~\the\figno}%
%\figinsert\figin{\centerline{#3}}\medskip\centerline{\vbox{\baselineskip12pt
%\advance\hsize by -1truein\noindent\footnotefont{\bf Fig.~\the\figno:} #2}}
%\bigskip\endinsert
\global\advance\figno by1}
\fi
\useblackboard
\message{If you do not have msbm (blackboard bold) fonts,}
\message{change the option at the top of the tex file.}
%\font\blackboard=msbm10 scaled \magstep1
\font\blackboard=msbm10
\font\blackboards=msbm7
\font\blackboardss=msbm5
\textfont\black=\blackboard
\scriptfont\black=\blackboards
\scriptscriptfont\black=\blackboardss
\def\Bbb#1{{\fam\black\relax#1}}
\else
\def\Bbb{\bf}
\fi
% *************************************
%
\def\yboxit#1#2{\vbox{\hrule height #1 \hbox{\vrule width #1
\vbox{#2}\vrule width #1 }\hrule height #1 }}
\def\fillbox#1{\hbox to #1{\vbox to #1{\vfil}\hfil}}
\def\ybox{{\lower 1.3pt \yboxit{0.4pt}{\fillbox{8pt}}\hskip-0.2pt}}

\def\mapr{\mathop{\longrightarrow}\limits}
\def\rightarrowbox#1#2{
  \setbox1=\hbox{\kern#1{${ #2}$}\kern#1}
  \,\vbox{\offinterlineskip\hbox to\wd1{\hfil\copy1\hfil}
    \kern 3pt\hbox to\wd1{\rightarrowfill}}}

\def\comments#1{}

\def\bz{\bar z}
\def\QC{\Bbb{C}}

\def\QZ{\Bbb{Z}}
\def\p{\partial}
\def\delbar{{\bar\partial}}

\def\half{{1\over 2}}
\def\Tr{{{\rm Tr~ }}}
\def\tr{{\rm tr\ }}

\def\ket#1{|#1\rangle}

\def\vev#1{\langle{#1}\rangle}

\def\CM{{\cal M}}
\def\CN{{\cal N}}
\def\CO{{\cal O}}

\def\CW{{\cal W}}

\def\mapr{\mathop{\longrightarrow}\limits}

\def\P{\BP}
\def\WP{\BW\BP}

\def\II{\relax{I\kern-.10em I}}

\def\pt{{pt}}

\def\IZ{\relax\ifmmode\mathchoice
{\hbox{\cmss Z\kern-.4em Z}}{\hbox{\cmss Z\kern-.4em Z}}
{\lower.9pt\hbox{\cmsss Z\kern-.4em Z}}
{\lower1.2pt\hbox{\cmsss Z\kern-.4em Z}}\else{\cmss Z\kern-.4em
Z}\fi}
\def\IB{\relax{\rm I\kern-.18em B}}
\def\IC{{\relax\hbox{$\inbar\kern-.3em{\rm C}$}}}
\def\ID{\relax{\rm I\kern-.18em D}}
\def\IE{\relax{\rm I\kern-.18em E}}
\def\IF{\relax{\rm I\kern-.18em F}}
\def\IG{\relax\hbox{$\inbar\kern-.3em{\rm G}$}}
\def\IGa{\relax\hbox{${\rm I}\kern-.18em\Gamma$}}
\def\IH{\relax{\rm I\kern-.18em H}}
\def\II{\relax{\rm I\kern-.18em I}}
\def\IK{\relax{\rm I\kern-.18em K}}
\def\IN{\relax{\rm I\kern-.18em N}}
\def\IP{\relax{\rm I\kern-.18em P}}
%\def\IX{\relax{\rm X\kern-.01em X}}
%this doesn't work
\def\IW{{\bf W}}

\def\inbar{\,\vrule height1.5ex width.4pt depth0pt}
\def\mod{{\rm\; mod\;}}

\def\p{\partial}

\font\cmss=cmss10 \font\cmsss=cmss10 at 7pt
\def\IR{\relax{\rm I\kern-.18em R}}

\def\End{{\rm End\ }}
\def\Hom{{\rm Hom}}
\def\Ext{{\rm Ext}}

\def\BR{\IR}
\def\BZ{\QZ} 
\def\BP{\IP}
\def\BW{\IW}
\def\BR{\IR}
\def\BC{\QC}

\def\lp10{l_P^{10}}
\def\lp11{l_P^{11}}
\def\R11{R_{11}}

\newbox\tmpbox\setbox\tmpbox\hbox{\abstractfont RUNHETC-2002-09}
\Title{\vbox{\baselineskip12pt\hbox to\wd\tmpbox{\hss
hep-th/0203173}\hbox{RUNHETC-2002-09}\hbox{\ \
IMSc/2002/03/01}\hbox{\ CPHT-RR035.0302}}}
{\vbox{
\centerline{D-branes on Calabi--Yau Manifolds and Superpotentials}}}
\smallskip
\centerline{
Michael R. Douglas$^{1,2,}$\footnote{$^{3}$}{
Louis Michel Professor}}
\smallskip
\centerline{Suresh Govindarajan$^{4}$}
\smallskip
\centerline{T. Jayaraman$^{5}$}
\smallskip
\centerline{Alessandro Tomasiello$^{6}$}
\bigskip
\bigskip
\medskip
\centerline{$^1$Department of Physics and Astronomy, Rutgers University,
 Piscataway, NJ 08855-0849 USA}
\medskip
\centerline{$^2$Isaac Newton Institute for Mathematical Sciences,
Cambridge, CB3 0EH, U.K.}
\medskip
\centerline{$^3$I.H.E.S., Le Bois-Marie, Bures-sur-Yvette, 91440 France}
\medskip
\centerline{$^{4}$Department of Physics, Indian Institute of Technology, Madras,
Chennai 600036 India}
\medskip
\centerline{$^{5}$The Institute of Mathematical Sciences, C.I.T. Campus,
Taramani, Chennai 600113 India}
\medskip
\centerline{$^6$Centre de Physique Th{\'e}orique, Ecole Polytechnique,
 91128 Palaiseau, France}  

\bigskip
\centerline{\tt Email: mrd@physics.rutgers.edu; suresh@chaos.iitm.ernet.in;}
\centerline{\tt jayaram@imsc.ernet.in; Alessandro.Tomasiello@cpht.polytechnique.fr}
\bigskip

\noindent
We show how to compute terms in an expansion of the world-volume
superpotential for fairly general D-branes on the quintic Calabi-Yau
using linear sigma model techniques, and show in examples that this
superpotential captures the geometry and obstruction theory of bundles
and sheaves on this Calabi-Yau.

\Date{March 2002}
\def\np{{\it Nucl. Phys.}}

%%%%%%%%%%%%%
\nref\agvafa{M. Aganagic and C. Vafa,
``Mirror Symmetry, D-Branes and Counting Holomorphic Discs,''
[arXiv:hep-th/0012041].}
\nref\aspdoug{
P.~S.~Aspinwall and M.~R.~Douglas,
``D-brane stability and monodromy,''
[arXiv:hep-th/0110071]. 
%%CITATION = HEP-TH 0110071;%%
}
%\nref\partres{C. Beasley, B. R. Greene, C. I. Lazaroiu, and 
%  M. R. Plesser, ``D3-branes on partial resolutions of abelian 
%  quotient singularities of Calabi--Yau threefolds'', {\it Nucl. Phys.}
%  {\bf B566} (2000) 599, hep-th/9907186.}
\nref\beil{A. A. Beilinson, ``Coherent sheaves on $\IP^n$ and
  problems of linear algebra'', {\it Funct. Anal. Appl.}
  {\bf 12} (1978) 214--216.
%%CITATION = NONE;%%
}
\nref\berdou{D. Berenstein and M. R. Douglas, work in progress.
%%CITATION = NONE;%%
}
\nref\botttu{R.~Bott and L.~W.~Tu,
{\it Differential Forms in Algebraic Topology,} 
Springer, 1982.
%%CITATION = NONE;%%
}
%\nref\bkr{T. Bridgeland, A. King, and M. Reid, ``Mukai implies McKay'', 
%  math.AG/9908027.}
\nref\bdlr{I. Brunner, M. R. Douglas, A. Lawrence, and C. R{\"o}melsberger,
  ``D-branes on the quintic,'', JHEP {\bf 0008}, 015 (2000)
[arXiv:hep-th/9906200].
%%CITATION = HEP-TH 9906200;%%
}
\nref\bs{I.~Brunner and V.~Schomerus,
``On superpotentials for D-branes in Gepner models,''
JHEP {\bf 0010} (2000) 016 [arXiv:hep-th/0008194]. 
%%CITATION = HEP-TH 0008194;%%
}
%\nref\twopar{P. Candelas, X. de la Ossa, A. Font, S. Katz, and 
%  D. R. Morrison,
%  ``Mirror symmetry for two parameter models -- I'', 
%  {\it Nucl. Phys.} {\bf B416} (1994) 481, [arXiv:hep-th/9308083].}
\nref\candelas{P. Candelas, X. C. de la Ossa, P. S. Green, and 
  L. Parkes, ``A pair of
  {C}alabi--{Y}au manifolds as an exactly soluble superconformal theory'',
  \np\ {\bf B359} (1991) 21.
%%CITATION = NUPHA,B359,21;%%
}
\nref\catfel{
A. S. Cattaneo and G. Felder,
``A path integral approach to the Kontsevich quantization formula,''
Commun.Math.Phys. 212 (2000) 591-611;
math.QA/9902090.
%%CITATION = MATH-QA 9902090;%%
}
%\nref\cox{D. Cox, ``Recent developments in toric geometry'', {\it Proc. Symp.
%  Pure Math.} {\bf 62:2} (1997) 389--436, alg-geom/9606016.}
\nref\dgr{F. Denef, B. R. Greene and M. Raugas,
``Split attractor flows and the spectrum of BPS D-branes on the Quintic,''
JHEP 0105 (2001) 012; hep-th/0101135.
%%CITATION = HEP-TH 0101135;%%
}
%\nref\inprogress{D.-E. Diaconescu and M. R. Douglas, work in progress.}
\nref\diacenh{D.-~E.~Diaconescu, ``Enhanced D-branes categories from 
string field theory,'' , JHEP {\bf 06} (2001) 016 [arXiv:hep-th/0104200].
%%CITATION = HEP-TH 0104200;%%
}
\nref\dd{D.-E. Diaconescu and M. R. Douglas, ``D-branes on Stringy
Calabi-Yau Manifolds,'' [arXiv:hep-th/0006224].
%%CITATION = HEP-TH 0006224;%%
}
%\nref\dg{D.-E. Diaconescu and J. Gomis, ``Fractional branes and boundary
%  states in orbifold theories,'' hep-th/9906242.}
%\nref\dr{D.-E. Diaconescu and C. R{\"o}melsberger, ``D-branes and 
%  bundles on elliptic fibrations'', {\it Nucl. Phys.} {\bf B574} (2000) 
%  245, hep-th/9910172.}%
\nref\Donaldson{S. K. Donaldson and R. P. Thomas,
``Gauge theory in higher dimensions,''
in: The Geometric Universe; Science, Geometry and the work of Roger Penrose,
eds. S. A. Huggest et al, Oxford Univ. Press 1998.
%%CITATION = NONE;%%
}

%\nref\dtopics{M. R. Douglas, ``Topics in D-geometry'',
%  {\it Class. Quant. Grav.} {\bf 17} (2000) 1057, hep-th/9910170.}
\nref\dougder{M. R. Douglas, ``D-branes, categories and {\cal N}=1 
supersymmetry,'' [arXiv:hep-th/0011017].
%%CITATION = HEP-TH 0011017;%%.
}
%\nref\pistable{M. R. Douglas, B. Fiol, and C. R{\"o}melsberger,
%  ``Stability and BPS branes'', hep-th/0002037.}
\nref\dfrtwo{M. R. Douglas, B. Fiol, and C. R{\"o}melsberger, 
  ``The spectrum of BPS branes on a noncompact Calabi--Yau'', 
 [arXiv:hep-th/0003263].
%%CITATION = HEP-TH 0003263;%%.
}
%%\nref\noncompact{M. R. Douglas, B. Fiol, and C. R{\"o}melsberger, 
%  ``The spectrum of BPS branes on a noncompact Calabi--Yau'', 
%  hep-th/0003263.}
\nref\dgm{M. R. Douglas, B. R. Greene, and D. R. Morrison, ``Orbifold
  resolution by D-branes,'' {\it Nucl. Phys.} {\bf B506} (1997) 84,
  [arXiv:hep-th/9704151].
%%CITATION = HEP-TH 9704151;%%.
}
\nref\dm{M. R. Douglas and G. Moore, 
  ``D-branes, quivers, and ALE instantons,''
  [arXiv:hep-th/9603167].
%%CITATION = HEP-TH 9603167;%%.
}
\nref\fmw{
R.~Friedman, J.~Morgan and E.~Witten,
``Vector bundles and F theory,''
Commun.\ Math.\ Phys.\  {\bf 187}, 679 (1997)
[arXiv:hep-th/9701162].
%%CITATION = HEP-TH 9701162;%%
}  
%\nref\fulton{W. Fulton, {\it Introduction to toric varieties}, 
%  Annals of Mathematics Studies, vol.~131, The William H. Roever Lectures in 
%  Geometry, Princeton University Press, 1993.}
%\nref\gepner{D. Gepner, ``Space-time supersymmetry in compactified
%  string theory and superconformal models," {\it Nucl. Phys.} 
%  {\bf B296} (1988) 757.}
%\nref\govj{S. Govindarajan and T. Jayaraman, 
%  ``On the Landau--Ginzburg description
%  of boundary CFTs and special Lagrangian submanifolds'', 
%  hep-th/0003242.}
\nref\govjs{S. Govindarajan, T. Jayaraman, and T. Sarkar, 
  ``World sheet approaches to D-branes 
  on supersymmetric cycles'', [arXiv:hep-th/9907131].}
\nref\gjs{
S.~Govindarajan, T.~Jayaraman and T.~Sarkar,
``On D-branes from gauged linear sigma models,''
Nucl.\ Phys.\ B {\bf 593}, 155 (2001)
[arXiv:hep-th/0007075].
%%CITATION = HEP-TH 0007075;%%
}
\nref\govjayhel{
S.~Govindarajan and T.~Jayaraman,
``D-branes, exceptional sheaves and quivers on Calabi-Yau manifolds:
 From Mukai to McKay,''
Nucl.\ Phys.\ B {\bf 600}, 457 (2001)
[arXiv:hep-th/0010196].
%%CITATION = HEP-TH 0010196;%%
}
\nref\gj{
S.~Govindarajan and T.~Jayaraman,
``Boundary fermions, coherent sheaves and D-branes on Calabi-Yau
manifolds,''
Nucl.\ Phys.\ B {\bf 618}, 50 (2001)
[arXiv:hep-th/0104126].
%%CITATION = HEP-TH 0104126;%%
}
\nref\gjsinst{
S.~Govindarajan, T.~Jayaraman and T.~Sarkar,
``Disc instantons in linear sigma models,''
[arXiv:hep-th/0108234]. 
%%CITATION = HEP-TH 0108234;%%
}  
\nref\GH{Ph.~Griffiths and J.~Harris, {\it Principles of algebraic geometry}, 
Wiley \& Sons, 1978.
%%CITATION = NONE;%%
}
\nref\hartshorne{R.~Hartshorne, {\it Algebraic geometry}, Springer-Verlag, 
1979.
%%CITATION = NONE;%%
}
\nref\hellmc{
S.~Hellerman and J.~McGreevy,
``Linear sigma model toolshed for D-brane physics,''
JHEP {\bf 0110}, 002 (2001)
[arXiv:hep-th/0104100].
%%CITATION = HEP-TH 0104100;%%
}
\nref\hellmck{
S.~Hellerman, S.~Kachru, A.~E.~Lawrence and J.~McGreevy,
``Linear sigma models for open strings,''
[arXiv:hep-th/0109069].
%%CITATION = HEP-TH 0109069;%%
}
\nref\horivafa{K.~Hori and C.~Vafa,``Mirror symmetry,''
[arXiv:hep-th/0002222].
%%CITATION = HEP-TH 0002222;%%.
}
\nref\hiv{K. Hori, A. Iqbal, and C. Vafa, 
  ``D-branes and mirror symmetry'', 
  [arXiv:hep-th/0005247].
%%CITATION = HEP-TH 0005247;%%.
}
%\nref\hulltown{C. M. Hull and P. K. Townsend, ``Enhanced gauge symmetries
%  in superstring theories'', {\it Nucl. Phys.} {\bf B451} (1995) 
%  525.}
%\nref\itonakajima{Y. Ito and H. Nakajima, ``McKay correspondence and
%  Hilbert schemes in dimension three'', math.AG/9803120.}
%\nref\johnsonmyers{C. V. Johnson and R. C. Meyers, ``Aspects of type 
%  IIB theory on ALE spaces'', {\it Phys. Rev.} {\bf D55} (1997) 6382, 
%  hep-th/9610140.}
%\nref\lerche{P. Kaste, W. Lerche, C. A. L{\"u}tken, and J. Walcher,
%  ``D-branes on K3 fibrations,'' hep-th/9912147.}
\nref\kac{V.~Kac, ``Infinite root systems, representations of graphs
and Invariant Theory,'' Invent.\ Mathem.\ {\bf 56}(1980) 57.
%%CITATION = NONE;%%
}
\nref\kontsoi{
M.~Kontsevich and Y.~Soibelman,
``Homological mirror symmetry and torus fibrations,''
[arXiv:math.sg/0011041].
%%CITATION = MATH-SG 0011041;%%
}
%%\nref\kontsevich{M.~Kontsevich, ``Homological algebra of mirror 
%%  symmetry'', alg-geom/9411018.}
%\nref\kronone{P. B. Kronheimer, ``The construction of ALE spaces as 
%  hyper-K{\"a}hler quotients'', {\it J. Differential Geom.} {\bf 29} (1989)
%  665.}
%\nref\krontwo{P. B. Kronheimer,
%  ``A Torelli-type theorem for gravitational instantons'',
%  {\it J. Differential Geom.} {\bf 29} (1989) 685.}
%\nref\kronnak{P. B. Kronheimer and H. Nakajima, 
%  ``Yang--Mills instantons on ALE gravitational instantons'', 
%  {\it Math. Ann.} {\bf 288} (1990) 263.}
\nref\lazainf{C.~I.~Lazaroiu, ``String field theory and brane 
superpotentials,'' JHEP {\bf 0110}, 018 (2001)
[arXiv:hep-th/0107162].
%%CITATION = HEP-TH 0107162;%%.
}
\nref\mayrhel{
P.~Mayr,
``Phases of supersymmetric D-branes on Kaehler manifolds and the McKay 
correspondence,''
JHEP {\bf 0101}, 018 (2001)
[arXiv:hep-th/0010223].
%%CITATION = HEP-TH 0010223;%%
}
\nref\mayr{
P.~Mayr, 
``N = 1 mirror symmetry and open/closed string duality,''
[arXiv:hep-th/0108229]. \semi
%%CITATION = HEP-TH 0108229;%%
W.~Lerche and P.~Mayr,
``On N = 1 mirror symmetry for open type II strings,''
[arXiv:hep-th/0111113].
%%CITATION = HEP-TH 0111113;%%
}
\nref\morrpless{
D.~R.~Morrison and M.~Ronen Plesser,
``Towards mirror symmetry as duality for two dimensional abelian gauge
theories,''    
Nucl.\ Phys.\ Proc.\ Suppl.\  {\bf 46}, 177 (1996)
[arXiv:hep-th/9508107].
%%CITATION = HEP-TH 9508107;%%
}
\nref\speseq{J.~Mac Cleary, {\it A user's guide to spectral sequences},
Cambridge, 2001.
%%CITATION = NONE;%%
}
%\nref\mckay{J. McKay, ``Graphs, singularities, and finite groups'', 
%  {\it Proc. Symp. in Pure Math.} {\bf 37} (1980) 183.}
\nref\merkulov{S.~Merkulov, ``Strong homotopy algebras of
a K{\"a}hler manifold,'' [arXiv:math.AG/9809172].
%%CITATION = MATH-AG 9809172; %%
}
\nref\mitchell{B.~Mitchell, ``Rings with several objects,'' {\it Adv. Math.}
8 (1972) 1-161.
%%CITATION = NONE;%%
}
%\nref\quantcoh{
%  D. R. Morrison and M. R. Plesser, 
%  `` Summing the instantons: Quantum cohomology and mirror symmetry 
%  in toric varieties'',  {\it Nucl. Phys.} {\bf B440} (1995) 279,
%  hep-th/9412236.}
%\nref\naknozak{M. Naka and M. Nozaki, 
%  ``Boundary states in Gepner models,'' {\it JHEP}\/ 0005 (2000) 027,
%  hep-th/0001037.}
\nref\polishchuk{A. Polishchuk, ``Homological Mirror Symmetry with
Higher Products,'' [arXiv:math.AG/9901025].
%%CITATION = MATH-AG 9901025; %%
}
\nref\rs{A. Recknagel and V. Schomerus, ``D-branes in Gepner models'',
  \np\ {\bf B531} (1998) 185, [arXiv:hep-th/9712186].
%%CITATION = HEP-TH 9712186;%%.
}
\nref\reid{M. Reid, ``McKay correspondence'', [arXiv:alg-geom/9702016].
%%CITATION = MATH-AG 9702016; %%
}
\nref\reidbourbaki{M. Reid, ``La correspondance de McKay'', S{\'e}minaire
  Bourbaki (novembre 1999), no. 867, 
  [arXiv:math.AG/9911165].
%%CITATION = MATH-AG 9911165; %%
}
%\nref\infirri{A. V. Sardo Infirri, ``Partial resolutions of orbifold
%  singularities via moduli spaces of HYM-type bundles,'' alg-geom/9610004.}
%\nref\infirritwo{A. V. Sardo Infirri, ``Resolutions of orbifold 
%  singularities and flows on the McKay quiver,'' alg-geom/9610005.}
%\nref\schei{E. Scheidegger, 
%  ``D-branes on some one- and two-parameter Calabi--Yau hypersurfaces,'' 
%  {\it JHEP}\/ 0004 (2000) 003, hep-th/9912188.}
\nref\seidel{P. Seidel, private communication.
%%CITATION = NONE;%%
}
\nref\seitho{P. Seidel and R. Thomas, ``Braid group actions on
derived categories of sheaves,'' [arXiv:math.AG/0001043].
%%CITATION = MATH-AG 0001043; %%
}
\nref\syz{
A.~Strominger, S.~T.~Yau and E.~Zaslow,
%``Mirror symmetry is T-duality,''
Nucl.\ Phys.\ B {\bf 479}, 243 (1996)  
[arXiv:hep-th/9606040].
%%CITATION = HEP-TH 9606040;%%
}  
\nref\tomainf{
A.~Tomasiello,
``A-infinity structure and superpotentials,''
JHEP {\bf 0109}, 030 (2001)
[arXiv:hep-th/0107195].
%%CITATION = HEP-TH 0107195;%%
}
\nref\tomhel{
A.~Tomasiello,
``D-branes on Calabi-Yau manifolds and helices,''
JHEP {\bf 0102}, 008 (2001)
[arXiv:hep-th/0010217].
%%CITATION = HEP-TH 0010217;%%
}
%\nref\warner{N.P. Warner,
%  ``Supersymmetry in Boundary Integrable Models,''
%  {\it Nucl. Phys.} B450 (1995) 663-694;
%hep-th/9506064.}
\nref\wittentop{E. Witten, ``Chern-Simons Gauge Theory as a String
Theory,'' 
[arXiv:hep-th/9207094].
%%CITATION = HEP-TH 9207094;%%
}
\nref\wittenphases{
E.~Witten,
``Phases of N = 2 theories in two dimensions,''
Nucl.\ Phys.\ B {\bf 403}, 159 (1993)
[arXiv:hep-th/9301042].
%%CITATION = HEP-TH 9301042;%%
}
%\nref\witten{E. Witten, ``Phases of $N{=}2$ theories in two dimensions,''
%  {\it Nucl. Phys.} {\bf B403} (1993) 159, hep-th/9301042.}
\nref\wittenqcd{E. Witten, ``Branes and the Dynamics of QCD'',
Nucl. Phys. {\bf B507}, 658 (1997) [arXiv:hep-th/9706109].
%%CITATION = HEP-TH 9301042;%%
}
%
%%%%%%%%%%%%%%%%%%%%%%%%%%%
%
\newsec{Introduction}

D-branes on Calabi-Yau manifolds provide a simple and general way to
embed $\CN=1$ supersymmetric gauge theories in string
compactifications, and are the starting point for many applications:
the general study of $\CN=1$ compactification and duality, model
building, geometric engineering of gauge theory, AdS/CFT with $\CN=1$
supersymmetry, and the study of BPS states.

Much of this work, starting with \dgm, has focused on D-branes on
noncompact Calabi-Yaus, and one starts to feel that that this case is
reasonably well understood.  The same ideas should apply just as well
to compact Calabi-Yaus, and this holds out the possibility of a much
more complete understanding of $\CN=1$ compactification than we have
now.

At present three approaches to this problem have shown some success.
First, for the special case of elliptically fibered Calabi-Yau, one
can do T-duality on the fiber, reducing much (but not all) of the
problem to the better understood theory of holomorphic curves and
surfaces in a CY \fmw.  Another approach is to study the theory of special
Lagrangian submanifolds on the mirror CY; this is also based on
T-duality, as has been explained in various ways \refs{\syz,\morrpless,\horivafa}.

The third approach grew out of the study of boundary states in Gepner
models \rs, which gave an explicit construction of certain D-branes in
a highly stringy and {\it a priori} nongeometric regime: in linear
sigma model terms, the Landau-Ginzburg (LG) phase.  Arguments from
topological open string theory, formalized in the ``decoupling
statement'' of \bdlr, suggest that each B type brane corresponds to a
specific holomorphic bundle on the CY in the large volume limit,
so the problem of identifying which bundle corresponds to
which boundary state is well-posed.

In \refs{\dfrtwo,\dd}\ it was found that these boundary states could
be largely understood in terms of a standard construction of sheaves
on projective space $\P^n$ formulated by Beilinson \beil.  Beilinson's
construction links algebraic and geometric data very tightly, and
seems to be the best general construction which emerged from much
mathematical work on the subject.  Following relevant mathematical
work on the generalized McKay correspondence \reid, this has been
generalized to subvarieties of more general toric varieties, in
\refs{\dd,\govjayhel,\mayrhel,\tomhel}.

Somewhat surprisingly, these constructions lead not just to sheaves
but to arbitrary objects in the derived category of coherent sheaves.
If one's goal is to study bundles, this is a disadvantage, as one must
then work to prove that the object of interest is a bundle.  However,
there is by now a lot of evidence that physical BPS branes are not
just bundles or even sheaves, but actually do correspond to objects in
the derived category \dougder.  The usual association with bundles and
sheaves is only appropriate in the large volume limit, while the
formalism of the derived category applies in the stringy regime as
well.  The physical question is not whether the objects of interest
are bundles, but whether they are stable.

To summarize what has been accomplished in this approach so far, we
can derive explicit $\CN=1$ supersymmetric field theories whose moduli
spaces are the classical (open string) moduli spaces of superstring
compactification on Calabi-Yau, given an important proviso: it must
turn out that sheaves on the ambient space, say $\P^n$, in which the
Calabi-Yau of interest is embedded, are the same or at least very
similar to sheaves on the embedded Calabi-Yau.  The usefulness of this
construction depends very much on this point.

At first sight one might be discouraged by examples in low dimension.
For example, one might try to study bundles on the torus $T^2$ by
realizing it as a curve embedded in $\P^2$.  In fact this would be a
very bad idea: the classification of bundles on $\P^2$ is very much
more complicated than that on $T^2$, and the moduli of bundles on
$T^2$ (for example of a flat connection on $T^2$) are typically not
even present before the restriction.

However, some more thinking about this point shows that this type of
construction gets better in higher dimensions, and becomes very useful
precisely when one reaches dimension three, i.e. the case of interest.
We will tackle the problem from two different but complementary
approaches, with roots in mathematics and physics.

Our first approach will use the mathematics of restriction of sheaves.
Thus, one first asks: when do two sheaves $E$ and $E'$ on the ambient space
restrict to the same sheaf on the CY? 
This can be studied
systematically using long exact sequences (or related constructions)
as we discuss in section 2 and 3, and one indeed finds that in higher
dimension this is generally not a difficult issue. 
One then asks, how many moduli are present for
D-brane configurations after restriction to the CY?  
The basic result here is, that on general grounds, linearized
moduli correspond to elements of $\Ext^1(E,E)$ and will arise as the
Serre duals of elements of $\Ext^{d-1}(E,E)$ on the ambient space;
again these are under control. 

One then needs to know when such linearized moduli correspond to true
moduli, i.e. whether they can be given finite vevs.  Mathematically
this is a complicated problem of obstruction theory, but it is here
that physics provides a big advantage: in principle the answer to all
such questions is encoded in the superpotential $\CW$ of the world-volume
$\CN=1$ theory; allowed finite values of the moduli are the ones which
satisfy the F-flatness conditions $\CW'=0$. 

Furthermore, one has techniques for computing the superpotential in
interesting models.  The leading term has been computed in both
the Gepner model \bs\ and the linear sigma model \dd;
it is cubic and agrees with
what is needed to reproduce the Beilinson construction.\foot{
See also \refs{\govjs,\hiv,\hellmck}\ for other work on D-branes 
in LG theories.}
This cubic superpotential can be understood from the geometry of
orbifold resolution, in the same way as for
the solved case of $\BC^3$
orbifolds.  However, while the superpotential
for the noncompact orbifold is exactly cubic, in the LG model
the world-sheet superpotential leads to higher order corrections.

In this work, we compute the leading such correction for the quintic,
and show how it fits with the geometric interpretation of B branes as
bundles on CY (or objects in the derived category).

In section 2, we discuss some examples on the quintic which demonstrate
the need for these corrections.  One could in principle compute them
by perturbation theory in holomorphic Chern-Simons theory;
we use this to give an expression for the next to
leading order term, at sixth order.  This expression is not completely
explicit, depending on complicated geometric data such as the Green
function for the Laplacian
on the CY, but the general form of the expression is clear.
We show that, given some plausible assumptions, this turns out to be
enough to determine the term.  We then show in examples that it plays
the anticipated role. 

In section 3, we discuss an alternate mathematical approach to getting
these corrections, using spectral sequences.

In section 4, we give background on the LG description  
and use it to compute the sixth order correction explicitly.
It takes precisely the form predicted by the geometric analysis.

In sections 5 and 6, we discuss a number of other examples,
further directions and connections with mathematics.

\newsec{Geometry of sheaves on the quintic}

As discussed in the introduction, the problem of describing the
holomorphic structure of B branes on the quintic can be approached
from several directions.  In this section we consider them as
coherent sheaves, or objects in the derived category of these.

\subsec{Sheaves on $\P^n$ and Beilinson's construction}

A broad class of CY threefolds arise as submanifolds of toric Fano manifolds.
This is good in simple cases such as the quintic, but becomes
cumbersome in general, when the codimension becomes high. Fortunately
the class stays large enough even restricting to the case of hypersurfaces.
This suggests that we understand sheaves on the CY as the restriction
of those from the ambient space.  This is standard procedure both in
physics and mathematics.  As we mentioned one can do it 
for more general toric ambient spaces, but here we just consider $\P^4$.

The simplest physical way to motivate the type of construction we use,
is that we will choose $n$ generating branes (a ``basis''), call them
$S_i$ with $1\le i\le n$.  We then get all branes by forming all
possible bound states of these.  This is done by considering the
world-volume theories of $N_i$ branes of type $S_i$; each BPS bound
state will be a supersymmetric vacuum of at least one of these
theories.

In the case of the quintic, one simple choice of basis is
the line bundles $\CO(n)$ for some sequence of five consecutive $n$.
These span the K theory of $\P^4$, and an index 25 sublattice of the K theory
of the quintic.  An equally good basis, dual in a sense explained in
\refs{\beil,\dd} is to take the
exterior powers of the cotangent bundle,
\eqn\sbasis{
S_{n}\equiv \wedge^{5-n} \Omega_{\P^4}(n-5)[n-1], \qquad 1\le n \le 5.  
}
We will denote the restriction of the $S_n$ to $\CM$ as $B_{n}$.
The $[n-1]$ will be explained later; in particular the terms with
$n-1$ odd are antibranes (with negative D$6$ charge).

Ideally, we would have a ``universal'' construction with the following
property: not only can all branes be constructed, each brane can be
constructed in only one way: this will mean that our world-volume theories
precisely reproduce the moduli space of branes.  This requires that the
multiplicities $N_i$ be determined by the K theory class, so clearly this 
requires us to take exactly $n=\dim H^{even}$ generating branes.

However, even in simple examples, it turns out that to form all stable
objects from such a basis, one must use both branes and the antibranes
of the generating set.  For this and many other reasons, one must work
in a formalism which can describe general brane-antibrane bound
states, while keeping track of all holomorphic information.  Such a
construction is the derived category of coherent sheaves.

There is a simple description of the derived category of coherent
sheaves on $\P^4$: it is the same as the derived category of quiver
representations of a supersymmetric gauge theory given in \dougder\
(and which we will review shortly).  There are various mathematical
and physical arguments that most, if not all, of these objects
(restricted to the quintic) can appear as BPS branes at different
points in K{\"a}hler moduli space.

\subsec{Examples on $\P^4$}
\subseclab\subsecex
We list here some simple examples which will reappear below.  
We will go through the translation from the geometry of sheaves 
to supersymmetric gauge theory in some detail in the first example.

The simplest example is a two term complex describing the bound state of
two elementary branes, say
\eqn\Eone{
\Omega(1) \mapr^f \CO \ .
}
The map $f$ is multiplication by a section of $(\Omega(1))^*= T(-1)$,
the tangent bundle twisted by $O(-1)$.
In physics notation, a section of $T(-1)$ is a tangent vector $f^i \p_i$,
specified by a set of functions $f^i$ (homogeneous of degree zero), 
modulo the ``gauge symmetry'' $f^i \sim f^i + z^i \epsilon$ for any
$\epsilon$ of degree $-1$.  
In mathematics, exactly the same idea is expressed by the Euler sequence
$$
0 \mapr \CO(-1) \mapr \IC^5 \otimes \CO \mapr 
T(-1) \mapr 0
$$
for which reason $T(-1)$ is often denoted by ${\cal Q}$ (for quotient). 

Similarly, a section of $\Omega(1)$ is a (degree zero) one-form
$\psi_i dz^i$ satisfying the constraint $z^i \psi_i = 0$.  The map
$f$ in \Eone\ is then $\psi_i \mapr f^i \psi_i$.

A holomorphic map of this type will have $f^i$ constant, so the space of these
is $\Hom(\Omega(1),\CO) \cong \BC^5$.  To realize this bound state of two
elementary branes in the supersymmetric gauge theory requires 
five chiral multiplets $X^i$ with charge $(-1,+1)$  
under $U(1)\times U(1)$; their
vevs specify a map as $f^i=X^i$.  The moduli space of supersymmetric vacua is
the space of these vevs modulo the off-diagonal $U(1)$ gauge fixing and D-term
constraint; if this constraint allows $X\ne 0$, the moduli space is $\P^4$.

The geometric interpretation of the bound state is the cohomology of the
complex \Eone.  
Such an $f$ has a kernel but no cokernel, so the cohomology of \Eone\ is a
sheaf of rank 3.  Call it $E_1$; 
we can express the construction with the exact sequence
$$
0 \mapr E_1 \mapr \Omega(1) \mapr^f \CO \mapr 0\ .
$$
The sheaf $E_1$ is singular at the point $z^i=X^i$, so the moduli space is just
the $\P^4$ parameterizing the choice of singular point.

In the derived category, this construction can be expressed as
$$
E_1 \cong \underline{\Omega(1)}\mapr \CO .
$$
As in \refs{\aspdoug}, we underline the zero position in a complex.

Another simple example is
\eqn\Etwo{
\Omega^4(4) \mapr^f \Omega^3(3) \mapr^g \Omega^2(2)\ .
}
The two maps $f$ and $g$ are again both multiplications by sections of
$T(-1)$, that is vectors in $\IC^5$. But the condition that \Etwo\ be a
complex (physically, as we will review later, this is the superpotential 
constraint) has as unique solution $X_f^i = X_g^i$
up to rescaling. These rescaling do not give different sheaves
(physically this is a gauge quotient), so these
objects also have moduli space $\P^4$.  This complex turns out to
have its cohomology at the last node (as one might guess since this
is the constituent of highest rank).  It is another rank 3 sheaf,
which we denote $E_2$.  In the derived category, $E_2$ is equivalent to
$\Omega^4(4) \mapr \Omega^3(3) \mapr \underline{\Omega^2(2)}$.

One can continue in this vein, eventually constructing
$$
\Omega^4(4) \mapr^X \Omega^3(3) \mapr^X
\Omega^2(2) \mapr^X \Omega(1) \mapr^X \CO .
$$
The constraints again force all the maps to be proportional, so all have
been denoted by their vectors $X$, and the moduli space is again $\P^4$.

This sequence turns out to have cohomology only at the last term and
this is in fact a Koszul resolution of a point on $\P^4$,
dual to the one with line bundles $\CO(k)$.  We denote
the structure sheaf of the point $z$ as $\CO_z$.

Using these facts, and exact sequences {\it of complexes}, 
we can interpret the sheaves introduced so far as fitting 
into the distinguished triangle in the derived category
\eqn\exactpoint{
E_1 \mapr \CO_z[-1] \mapr E_2[1] \mapr E_1[1]
}
So $E_1$ and $E_2$ are very similar, but differ by the addition of
a point. Physically speaking, this implies that $E_1$ and $E_2$ differ
by the addition of some $0$-branes on $\P^4$.

Constructing more general sheaves will of course require taking some
of the $N_i$ different from $0$ or $1$.  Here are three examples.

First, we can ask how the other more familiar objects on CY such as
curves, hypersurfaces etc. look in this picture.  Most of these
will be intersections of objects of one dimension higher with the
defining equation of the CY.  Objects with nonzero codimension on
$\P^4$ will be realized in the basis of line bundles by their Koszul
resolutions; these can then be ``dualized'' to obtain an alternative 
Koszul resolution in the style of what we saw above for a point $z$.
Alternatively we could simply plug the initial sheaf in the Beilinson 
machinery and obtain the same expression.

Consider the structure sheaf $\CO_P$ of a hyperplane $P$ in $\P^4$.
Slightly simpler for our purposes is $\CO_P(1)$, its
twist by $\CO(1)$.  In terms of line bundles this is
$$
\CO_P(1) \cong \CO \mapr^f \underline{\CO(1)}.
$$
In the Beilinson basis, it is
\eqn\OPone{
\CO_P(1) \cong \Omega(1) \mapr^f \underline{\CO \otimes \BC^4} .
}
Here $f$ is a $1\times 4$ matrix of the the same maps used in \Eone;
this leads to a moduli space of dimension $20-16=4$.  By definition,
this moduli space is the dual $\P^4$. 

We can construct this moduli space in a more explicit way as follows. 
The map $f$ in \OPone\ 
is again in the form $X^i$, but each component is now a
matrix,
$$
\Hom( \CO(1), \CO_P(1)) \mapr^{X^i} \Hom( \CO, \CO_P(1))\ .
$$
The generator of the first group is simply restriction on $P$; its image
in the second group is simply $(e_i)_{|_P}$, where $e_i$ is one of the
basis elements of $\Hom(\CO, \CO(1))=H^0(\CO(1))$. In particular, if
for example $P$ is described by $z_1=0$, $X_1$ is the zero map (being 
its image $(e_1)_{|_P}=0$. So we have 5 vectors in $\IC^4$;
or 4 vectors in $\IC^5$. These 4 vectors span exactly the hyperplane, 
and we recover the moduli space as $\IP^4$.
 
Subvarieties of higher codimension are constructed as longer complexes.
For example, consider a surface $\Sigma$ in $\P^4$ defined as the
intersection of two hyperplanes.
One can easily match the K theory-class in the Beilinson basis, giving
$$
\Omega^2(2) \mapr^f \Omega(1) \otimes \BC^3 \mapr^g \CO \otimes \BC^6 
  \mapr \CO_\Sigma(2)\ .
$$
An instructive exercise is to compute the tensor product of the
structure sheaves of two hyperplanes, $\CO_P$ and $\CO_Q$.  Doing this
in the line bundle basis leads to the Koszul complex; the complex above
is dual to this.

As another example, one of the Recknagel-Schomerus states
$\ket{11000}$ corresponds to
\eqn\nextex{
\Omega^2(2) \mapr^X \Omega(1) \otimes \BC^2 \mapr^X \CO\ .
}
From conformal field theory, it is easy to see
that this state has $11$ linearized moduli.

\subsec{Restriction to a hypersurface}
\subseclab\subsechyp

We now consider a degree $d$ hypersurface $\CM$ in $\P^{D+1}$ 
($\CM$ is CY if $d=D + 2$),
and explain why in complex dimension $D\ge 3$,
sheaves on a hypersurface are very closely related to those on the
ambient space, in distinction to the lower dimensional cases.
We start with the question, can two different sheaves $E$ and $E'$
restrict to the same sheaf on $\CM$?

Let us start with the simplest case of restricting a bundle.
Geometrically, one pictures a vector bundle as locally trivial over
patches $\alpha$, related by transition functions $g_{\alpha\beta}$, with 
two different bundles equivalent if they are related by 
holomorphic gauge transformations
$g'_{\alpha\beta} = g_\alpha g_{\alpha\beta} g_\beta^{-1}$.
In this language, the question is whether allowing gauge transformations
$g_\alpha$ which can become singular off $\CM$ leads to additional
equivalences between bundles.

This question can be rephrased more mathematically as follows:
does there exist an isomorphism between $E$ and $E'$ on $\CM$, i.e. an
invertible $\Hom(E,E')$, which is not just the restriction of a
$\Hom(E,E')$ between the bundles on the ambient space?

There is an exact sequence which is relevant for this question, which
relates the cohomology of bundles on $\CM$ to that on the ambient space.
It can be derived by tensoring the bundles on the ambient space with
the exact sequence
$$
0 \mapr \CO_{\P^{D+1}}(-d) \mapr \CO_{\P^{D+1}} \mapr \CO_\CM \mapr 0 .
$$
%Flat iff locally free! 
We obtain
\eqn\basseq{\eqalign{
0 \mapr &\Hom(E,E'(-d)) \mapr \Hom(E,E') \mapr \Hom(E,E'|_\CM) \mapr \cr
        &\Ext^1(E,E'(-d)) \mapr \Ext^1(E,E') \mapr \Ext^1(E,E'|_\CM) \mapr \cr
        &\hskip 4cm\ldots \cr
        &\Ext^{D+1}(E,E'(-d)) \mapr \Ext^{D+1}(E,E')
                \mapr \Ext^{D+1}(E,E'|_\CM) \mapr0\ .
}}
Here $\Hom\equiv\Hom_{\P^{D+1}}$ and $\Ext\equiv\Ext_{\P^{D+1}}$ are
on $\P^{D+1}$. The last groups in each row can be rewritten using
 the following basic fact
\hartshorne: $\Ext^i_M(E,{\cal E}) = H^i(M, E^*\otimes {\cal E})$ whenever
$E$ is a bundle on $M$ and ${\cal E}$ whatever sheaf of $\CO$-modules. 
In our case we have
$$\eqalign{
\Ext^i(E,E'|_\CM)\cong H^i(\P^{D+1}, E^* \otimes E'|_\CM)\cong&
H^i(\P^{D+1}, (E^* \otimes E')|_\CM)\cr
& \cong H^i(\CM, (E^* \otimes E')|_\CM)\cong
\Ext^i_\CM(E|_\CM,E'|_\CM)\ .}
$$
We also have Serre duality on $\P^{D+1}$, 
which states that
$$
\Ext^{D+1-n}(E,E') \cong \Ext^n(E',E(-D-2))^*
$$
and allows rewriting first groups in each row.
Summing up, we can rewrite \basseq\ as 
\eqn\enhbasseq{\eqalign{
0 \mapr &\Ext^{D+1}(E',E(d-D-2)) 
         \mapr \Hom(E,E') \mapr \Hom_\CM(E|_\CM\,E'|_\CM) \mapr \cr
        &\Ext^D(E',E(d-D-2)) 
         \mapr \Ext^1(E,E') \mapr \Ext^1_\CM(E|_\CM,E'|_\CM) \mapr \cr
        &\hskip 4cm \ldots \cr 
        &\Hom(E',E(d-D-2)) \mapr \Ext^{D+1}(E,E') \mapr0\mapr0\ .
}}

From this we can see that any $\Ext^i_\CM$ will receive contributions 
from both $\Ext^i_{\P^{D+1}}$ and a $\Ext^{D-i}_{\P^{D+1}}$ twisted by 
$d-D-2$.\foot{
A similar story can be told about the more general case in which 
we start from a Fano variety and a section of the anticanonical bundle.
This also works well if this bundle is ample, which will be the case
if the original weighted projective space did not have singularities
intersecting the hypersurface.  On the other hand, ampleness can be violated
if we have to resolve singularities which intersect the hypersurface.
A simple example is a hypersurface in $\WP_{1,1,1,3,3}$.
Physically, the rational boundary states in these Gepner models do not
span the K theory.
})
For the case of $\CM$ a CY (i.e. $d=D+2$), we see that extra
$\Ext^i(E,E')$ will come from $\Ext^{D-i}(E',E)$ on $\P^{D+1}$, or
$H^{D-i}(M,E'^* \otimes E)$ if $E'$ is a bundle.

Now for $[E]=[E']$ (both have the same K theory class), $c_1(E'^*
\otimes E)=0$.  For such bundles on a Fano variety, typically
the higher cohomology groups vanish for
sufficiently high degree, as formalized in vanishing theorems \GH.
We will make this more precise below in the case of most interest to us.
In general, it is easy to come up with examples with nonzero $H^1$ (any object
with moduli), and examples are known with nonzero $H^2$ (any object
with obstructed deformations), but already $H^3$ is not typical.

In $D=1$, this means that one always gains extra linearized moduli (elements
of $\Ext^1(E,E)$) upon restriction, and typically gains extra elements of
$\Hom(E,E')$ as well.  This leads to drastic differences between bundles on
the ambient space and the hypersurface.

In $D=2$, every $\Ext^1(E,E)$ on $\P^3$ will lead to a Serre dual
$\Ext^1(E,E)$ on K3 by this argument.  This type of pairing is
required by general considerations (we know that moduli spaces of
bundles on K3 are hyperk{\"a}hler) but again complicates the story.

In $D\ge 3$, the new morphisms required by Serre duality on $\CM$ are
higher cohomology, so they do not affect the story as directly.  The
pairing $\Ext^1 \cong \Ext^{2*}$ in $D=3$ is quite important as it
is responsible for the fact that the space of holomorphic objects can
be described by a superpotential, as we discuss below.

\subsec{Quiver gauge theory and spherical bundles} 

The mathematical considerations of the previous subsection have their
most direct physical translation in the quiver gauge theories of
B-type branes.
The relevant quiver gauge theory on $\IP^4$, or more precisely the
quiver gauge theory for the non-compact orbifold CY, $\IC^5/\BZ_5$, 
was written down in \dd. 
The nodes of the quiver correspond to the basis branes $S_i$,
and links to matter fields $\Ext^p(S_i,S_j)$.

We review the LG orbifold derivation of the quiver theory in section 4,
but here summarize the result.  The spectrum contains all odd $p$-forms
on $\BC^5$, 
i.e. $p=1,3,5$, and the orbifold projection enforces
$j-i=p (\mod 5)$.  We denote the corresponding bifundamental matter
as $X^i$ with $1\le i\le 5$ (for $p=1$), and
$Y^{ijk}$ (for $p=3$; we also write $Y_{ij}=\epsilon_{ijklm}Y^{klm}$).
There is also a $p=5$-form $Z^{ijklm}$ whose bosonic component has
string-scale mass and which will not play a role until later, so we
drop it here.  The resulting quiver with
$X$ arrows in groups of five and $Y$ arrows in groups of
ten is
\midinsert
\centerline{\epsfxsize=2.5in\epsfbox{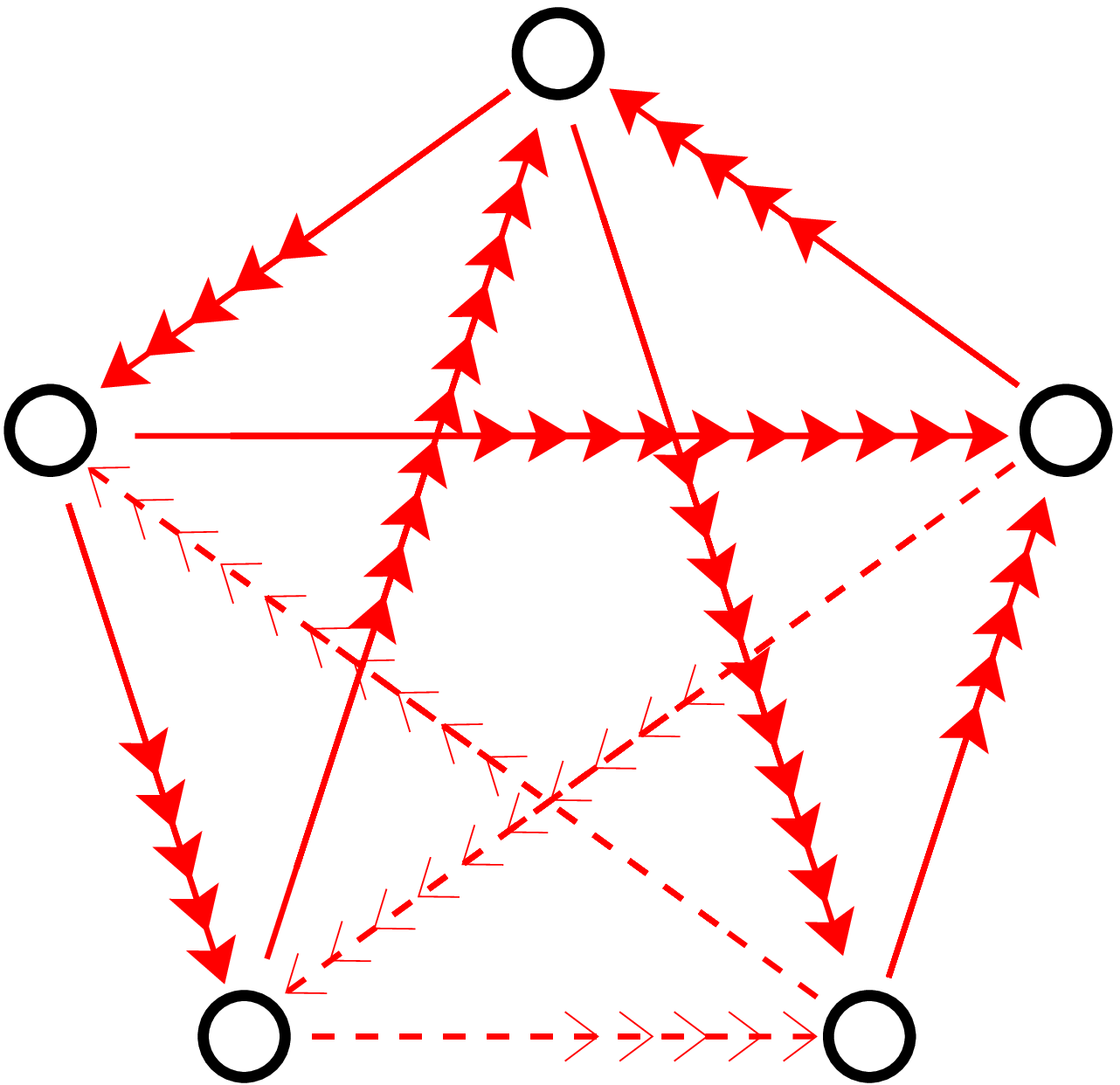}}
\centerline{Figure 1. The quintic quiver.}
\endinsert

Following the McKay correspondence, we make the identification \sbasis.
As explained in \dougder, one should include ``flow of gradings'' 
$[n-1]$ in comparing these objects to the large volume limit.
The simplest argument for this is that the large volume interpretation of the
fields $X^i$ is as the holomorphic maps
$\Hom(S_i, S_{i+1})$ (exactly the ones which appeared in \Eone),
which between branes and antibranes lead to tachyonic matter multiplets.
However, at the orbifold point, the $S_i$ are all ``branes'' and the
$p=1$ matter $X^i$ is massless.  These two assignments of
gradings are related precisely by the flow $S_i\rightarrow S_i[i+1]$.  
Similarly the $Y^{[ijk]}$ come from $\Hom(S_i,S_{i+3})$.
Note that the solid lines and dotted lines in figure 1 behave 
differently in the large volume limit; the dotted lines are obtained
by flow from the Serre duals of the maps above.

We also have the cubic superpotential
$$
\CW= X^i X^j Y_{ij} .
$$
We will review the LG computation of this in section 4.
It can also
be understood as a five-form analog of the holomorphic Chern-Simons action.

Now, solutions of $\CW'=0$ (mod complex gauge equivalence) are exactly
objects which can be used in a Beilinson construction of the derived
category of sheaves on $\BC^5/\BZ_5$, in direct analogy to the
familiar case of $\BC^3/\BZ_3$.  All moduli of these sheaves are
explicitly present; in gauge theory terms, as allowed variations
of the fields $X$ and $Y$.
In mathematical terms, these are the general 
deformations of a complex, meaning anything with total
degree 1, $H^j(\CM,S_i\otimes S_{i+k})$, $j+k=1$; some of this
is described in \dougder\ and in appendix A.

We now want to study the restriction of this entire category to the
quintic hypersurface.  We start by restricting the basis branes,
$B_i = S_i|_{\CM}$.  Since the $S_i$ have no
higher cohomology $\Ext^p(E,E)$ for any $p\ge 1$,
one can see using the long exact sequence \enhbasseq\ that all of these
restrict to ``spherical bundles'' on $\CM$, i.e. bundles with
$\Hom(E,E)=\Ext^D(E,E)=\BC$ and no other cohomology.

Thus the nodes of the quiver gauge theory on $\IP^4$ acquire no new
adjoint matter on restriction.  Similarly, one can check that the 
matter content $\Ext^p(B_i,B_j)$ agrees with the $\BC^5/\BZ_5$ quiver,
but with different values of $p$ in the cases where Serre duality was
involved, because one is now doing this in three dimensions.
The flow of gradings also works in a similar way, but now with shifts
$[3(n-1)/5]$ instead of $[n-1]$, as shown in \dougder.

So far making the restriction has been rather trivial, supporting the
rather optimistic claims of the previous subsection.  However, the
most obvious sign that not everything can be this simple is that the
K theory of the quintic is a rank $4$ group, while we have five
distinct elementary branes $B_i$.  This is the appropriate number for
$\P^4$, but it is clear that a given K theory class on the quintic can
be represented as a bound state in this basis in more than one way.
The basic example of this is the point on $\P^4$, which realizes the
fourth Chern class $c_4$.  A generic point on $\P^4$, not lying on the
hypersurface, restricts to nothing on the hypersurface, consistent
with losing this Chern class.\foot{
A point on the hypersurface restricts to a two-term complex
$O_{\pt} \mapr^0 O_{\pt}$, a direct sum of the $D0$ and $\bar D0$.
This is unstable and physically would annihilate to the vacuum, but
as a holomorphic object it is distinct from the null object.}

Thus, there must be equivalences between
configurations which naively are distinct.
We can get a simple example by considering the 
distinguished triangle \exactpoint.\foot{
This example includes as a special case the equivalence between the
Recknagel-Schomerus boundary states $\ket{10000}$ and $\ket{20000}$ following
from CFT field identifications.  We thank J. Walcher for this comment.}
Since a generic point on $\P^4$ restricts to nothing,
this directly implies that
\eqn\equival{
(E_1)|_\CM \cong (E_2)|_\CM .
}

Now this equivalence holds within the derived category of the quiver theory,
and seeing it directly within supersymmetric gauge theory might not be easy
(it is similar to the problem of describing brane-antibrane annihilation).
However, a simple prediction that can be checked is that $E_1$
and $E_2$ should come in the same moduli spaces.  
On the other hand, according
to quiver theory with the cubic superpotential,
$E_2$ has more than $4$ moduli on $\CM$, thanks to the extra fields $Y$.

Evidently this quiver description of bundles on CY is flawed. 

\subsec{Higher order superpotential terms from holomorphic Chern-Simons}
\subseclab\merone

It does not take very much searching to find the flaw.
From a physical point of view, there is no reason to think the cubic
superpotential we postulated for $\P^4$ is correct for branes on
$\CM$; it could have higher order corrections.  We will discuss the
linear sigma model computation of these corrections below.

These higher order corrections can be understood geometrically as
well.  In this language, the superpotential is essentially the
holomorphic Chern-Simons action
\eqn\hcs{
S= \int\Omega\wedge(\half A\bar\partial A + {1\over3} A\wedge A \wedge A)\ .
}
Naively, this action is cubic in
variations of the gauge field.  However, if one studies higher order
deformations, one must do perturbation theory, or its equivalent.  To
express the superpotential in terms of linearized deformations, one
must integrate out massive fields, which are not holomorphic, and this
will produce higher order terms in the superpotential.

This phenomenon was described in \wittentop\ and was previously worked
out in a more abstract context by various mathematicians; in the
present context it seems to appear first in Merkulov \merkulov.  
It is also discussed in \refs{\polishchuk,\tomainf,\lazainf,\kontsoi}.

Let us now see more explicitly how these corrections arise from this point
of view.  We can describe these corrections as arising from 
Feynman diagrams \refs{\lazainf,\tomainf} of the holomorphic Chern-Simons theory
\hcs.  Every boundary operator, including every $\Ext^1$ corresponding to
a linearized deformation, is an external leg.  The vertex is the
exterior product on forms with the tensor product of bundle factors.
Finally, the propagators
are the inverses of $\delbar$ operators of the appropriate bundles. (These 
inverses can be defined more properly using Hodge theory as 
$\delbar^{-1}=G\delbar^\dagger$; zero modes do not propagate.) 
So the cubic term $X^i X^j Y_{ij}$ 
we referred to earlier comes trivially from the vertex 

\midinsert
\centerline{\xy <0.4cm,0cm>:
(0,0)*+{X}; (1.732,1)="a"; **@{-};
(3.464,0)*+{X}; "a"; **@{-};
"a"; (1.732,3)*+{Y}; **@{-};
\endxy\ .}
\centerline{Figure 2. The cubic contribution to the superpotential.}
\endinsert

In general, higher point functions involve a sum over
planar tree diagrams of $\phi^3$ field theory.
Mathematically, this has an interpretation in
terms of $A_\infty$ algebras and categories.

Let us restrict attention to a quiver with three nodes, which can
describe the object $E_2$.  In this case, fortunately, there is only a
single diagram.  To see this, start by noting that only operator
orderings which lead to gauge-invariant world-volume couplings can be
non-zero; thus the leading correction will take the form
$\Tr YXXYXX$.
These operators correspond to zero-forms $X^i$, and three-forms (more
precisely $(0,3)$ forms)
$Y_{ij}$. Now we cannot have a vertex with $Y$ replaced
by a propagator: this would be $\delbar^{-1}(XX)$, and we don't have any
$(-1)$-form. We can instead write the two-form 
$\omega_2\equiv\delbar^{-1}(YX)$,
the one-form $\omega_1 \equiv\delbar^{-1}(X \delbar^{-1}(YX))$,
or a function $\omega_0$.  
If we want to make a $(0,3)$ 
form which then we can integrate against $\Omega$, the only 
options are $\omega_3 XY$, or $\omega_2 \omega_1$. These two 
choices are equivalent by integration by parts: 
$$
\int \Omega YX \delbar^{-1}( X \delbar^{-1}(XY) )=
\int \Omega \delbar\,\delbar^{-1}(YX) \delbar^{-1}( X \delbar^{-1}(XY) )
= -\int \Omega \delbar^{-1}(YX)  X \delbar^{-1}(YX) \ .
$$

The final result for this term in the superpotential corresponds to the
planar Feynman diagram in figure 3.
Proceeding further with this computation would require
us to have a concrete representation of
$\delbar^{-1}= G \delbar^\dagger$.  We do not, and instead we will
compute this term in section 4 with topological field theory and
linear sigma model methods.

\midinsert
\centerline{\xy <0.5cm,0cm>:
(0,0)*+{X}; (1,1)="a"; **@{-};
(0,2)*+{Y}; "a"; **@{-};
"a"; (4,1)="b"; **@{-};
(5,0)*+{X}; "b"; **@{-};
(5,2)*+{Y}; "b"; **@{-};
(2,1);(1.66,2)*+{X};**@{-};
(3,1);(3.33,2)*+{X};**@{-};
\endxy\ .}
\centerline{Figure 3. The sextic contribution.}
\endinsert

\subsec{Higher order superpotential terms resolve the problem}

Although the expression we just derived is not too explicit, it
turns out to be good enough, in the sense that with some further physical
input we can write a unique corresponding term in the superpotential.

The computation we described leads to a term
\eqn\tenthorder{
I^{(6)}_{j_1j_2j_3j_4j_5j_6k_1k_2k_3k_4}
\Tr Y^{j_1j_2j_3} X^{k_1} X^{k_2} Y^{j_4j_5j_6} X^{k_3} X^{k_4} ,
}
where $I^{(6)}$ is some tenth order tensor.
This tensor is constrained by symmetry; for example it is antisymmetric
in $j_1j_2j_3$, and so forth. 

Let us write an example of a tensor with the correct symmetries:
$$
I^{(6)}_{j_1j_2j_3j_4j_5j_6k_1k_2k_3k_4} =
 \epsilon_{j_1j_2j_3j_4j_5} c_{j_6k_1k_2k_3k_4} .
$$
where $c_{ijklm}$ is the totally symmetric tensor defining the
polynomial which describes the quintic, $c_{ijklm}z^i z^j z^k z^l z^m=0$.

Of course there are many other ways we could have contracted the indices.
However, if we are interested in solutions of $W'=0$, it suffices to know
the sixth order term up to corrections proportional to $W'$ at the
lower order.  In other words, we can assume the relations
$$X^i X^j = X^j X^i$$
and
$$X^i Y_{ij} = Y_{ij} X^j = 0.$$
This allows fairly general reorderings of the indices, and one
can show that all of the possible invariants constructed from the
two tensors $\epsilon$ and $c$ are equivalent up to these relations.

From the point of view of deformations, we must start with 
initial data satisfying the first relation, and the second relation 
survives the deformation. We discuss
this further in the next section.

Let us check that this term resolves our contradiction,
while not spoiling other good results.
We start from the example of $E_1$ and $E_2$ from above, that initially 
prompted our discussion.  We start with
a solution for the $E_2$ quiver theory with $Y=0$.  This requires all
$X$ equal; by global symmetry we can take them to be
$X^i=\delta^{i,1}$.  We should then consider hypersurfaces which do not
intersect the point $z^i=X^i$ (otherwise $O_z$ will not restrict to
zero and in fact $E_1\ne E_2$).  This requires $c_{11111}\ne 0$.

With the additional superpotential term, the equations $W'=0$, linearized
about this configuration as we saw above, become explicitly
\eqn\lindef{
\eqalign{
& 0 = \epsilon_{ijkl1} y^{ijk} \qquad \qquad \qquad \qquad \qquad \qquad
(X \cdot y = 0 = y \cdot X')\cr
&\matrix{\ 0 = x^i + 
 c_{1111j} y^{ij1} \qquad \forall i\ne 1 \cr
0 = c_{1111k} y^{ijk} \qquad \forall i,j\ne 1}\qquad \qquad 
(x'\wedge X + X' \wedge x + X^2 X'^2 y=0) 
}}
(remember that lower case letters denote candidate deformations; we
drop the subscript $_0$ on initial-point values; the quintic
term is now given by the superpotential we just computed).
The first equation sets to zero all components of $y^{ijk}$ with
$i,j,k\ne 1$; this leaves us with only the $y_{1jk}$, $j,k \ne 1$, which are
our 6 candidate moduli.  The third equation sets all of the other 
components to zero, if $c_{11111}\ne 0$: this means that there are no
deformations arising from the $y_{ijk}$. 

Thus both objects have the same moduli space and the contradiction is
eliminated.  The basic lesson is that the effect of restriction is in
general non-trivial, but can be summarized in the higher order terms
of the world-volume superpotential.  We will illustrate this with more
examples in section 5.

\newsec{Restriction using spectral sequences}
\seclab\thedouble

An alternate approach to studying the restriction of sheaves is to use
a spectral sequence, as we discuss in this section.  This is not really
required for the rest of the discussion, but is also instructive.

The phenomena we saw above would be visible already in the linearized
deformations, if we can find these around a general complex.  In this
special case, one can get the correct deformation theory as a
cohomology problem, but now using a cohomology which includes the
nonholomorphic maps.  The potential relevance of this cohomology was
suggested by Diaconescu \diacenh.

In terms of topological open string theory, 
what we will do now amounts to
defining $Q_{E,F}=\delbar + Q_E - Q_F$ and allow for
possible non-holomorphic maps; in other words we will consider a
$n$-form $\psi^{(n)}$ to be $Q-exact$ if it obeys an equation like 
$Q \psi^{(n)} = \delbar \psi^{(n-1)}$.
This looks complicated, and it is, but the complications can
be reduced by appealing to the machinery of spectral sequences.

We begin with the special case of a complex
\eqn\example{
C \equiv \ \{ B_1 \mapr^{a_1} B_2 \mapr^{a_2} B_3 \}\ .
}
Starting with this complex, we proceed to construct a double complex.
This is needed because deformations of a single holomorphic bundle $B$ 
are computed from $H^1$ of a complex, for example the Dolbeault one:
\eqn\cohomcpl{
\Omega^{0,0}(\CM, B) \mapr \Omega^{0,1}(\CM, B) \mapr \ldots
}
where $\Omega^{p,q}(\CM, B)$ are forms of type $(p,q)$ with 
values in the bundle
$B$, and the maps in the complex are the Dolbeault differentials 
$\bar\partial_B$. The first cohomology of this complex 
$H^{0,1}_{\bar\partial}(\CM, B)$ can be indeed reexpressed by 
Dolbeault theorem in the more familiar form $H^1(\CM, B)$.
The fact that we are dealing with a complex of bundles gives
us another ``direction'' for the complex computing deformations. More
precisely, we have to write down the double complex $E^0_{p,q}$ 
whose $p$-th line 
($p=0,\ldots,3$) is, for our three-term example
\eqn\linedc{
\eqalign{\Omega^{0,p}(\CM, B_3^*\otimes B_1 ) \ 
\mapr^{\tilde d'} \ & 
\Omega^{0,p}(\CM, \oplus_{i=1}^2 \, B_{i+1}^*\otimes B_i ) \ 
\mapr^{\tilde d} \    
\Omega^{0,p}(\CM, \oplus_{i=1}^3 \, B_i^*\otimes B_i )   \cr
&\ \mapr^d \ 
\Omega^{0,p}(\CM, \oplus_{i=1}^2 \, B_i^*\otimes B_{i+1} ) \ 
\mapr^{d'}\  
\Omega^{0,p}(\CM, B_1^*\otimes B_3 ) \ .} 
}
The maps in these lines are
defined in a way which is similar to the complex one uses to compute $\Hom$ 
and $\Ext$ for a quiver without relations. For our three-term example, for
instance, first map $\tilde d'$ 
in \linedc\ is $f \mapsto (f\circ a_2, - a_1 \circ f)$.
Tildes are there to emphasize how these maps are duals among them; see later. 
Names of these maps will occasionally mean both the maps between the bundles
and, as in this case, the maps on the bundle-valued forms that they induce.
First cohomology group of the double complex defined in this way gives then
deformations of the original complex \example. 

To compute the cohomology of a double complex (or at least a good 
approximation to it: we will come back to this later),
the tool which is usually effective
is a spectral sequence. For a general introduction, see for instance 
\refs{\botttu,\GH,\speseq}: here we will 
concretely follow the procedure and explain it. Spectral sequence procedure
computes for us a series of approximations to the sought-for cohomology.
First approximation we choose to be the cohomology of the vertical arrows.
This means that we can replace each column (which has the form of \cohomcpl) 
with its cohomology. This gives us a new double complex, whose $p$-th line
is now 
\eqn\linedcf{
\eqalign{
H^{0,p}(\CM, B_3^*\otimes B_1 ) \ \mapr \ &
H^{0,p}(\CM, \oplus_{i=1}^2 \, B_{i+1}^*\otimes B_i ) \ \mapr \    
H^{0,p}(\CM, \oplus_{i=1}^3 \, B_i^*\otimes B_i ) \cr
&\ \mapr \ H^{0,p}(\CM, \oplus_{i=1}^2 \, B_i^*\otimes B_{i+1} ) \ \mapr \  
H^{0,p}(\CM, B_1^*\otimes B_3 ) \ , }
}
where the horizontal maps are the maps induced on the cohomologies 
by the horizontal maps in \linedc.
This is our first approximation to the cohomology of the double complex. 

So far we never specified what the $B_i$ were; we now
choose them to be $S_i|_{\CM}=(\Omega^{5-i}(5-i))|_\CM$ as in \sbasis.
Then the  cohomology groups in \linedcf\ are 
the $\Ext^p_\CM$ of the $B_i=(S_i)|_{\CM}$ 
and we can compute them from the $\Ext^p_{\IP^4}(S_i, S_j)$ again using
\enhbasseq. The result is that the only nonzero groups are
$$
\Hom_\CM(B_i,B_j)= \Hom_{\IP^4}(S_i,S_j)=\Lambda^{j-i}\,V =
\Ext^3_\CM(B_j,B_i)\ , \quad i\leq j \ ,
$$
where $V\equiv \IC^5$. More generally we can say that 
an exceptional series on $F$ becomes a so-called {\it spherical} series
of bundles on $\CM$ \seitho, 
analogously to what stated above about a single bundle. 
The double complex is thus now
\eqn\dcf{ E^1_{p,q}\equiv
\matrix{\xymatrix{\Lambda^2 V \ar[r]^{H(\tilde d')} & 
V \oplus V \ar[r]^{H(\tilde d)} & \IC^3\\
&& 0 \ar[u] \\ && 0 \ar[u] \\
&&\IC^3 \ar[u]\ar[r]^{H(d)} & V \oplus V \ar[r]^{H(d')} & \Lambda^2 V }}
%\diagram[Postscript=dvips]
%&\Lambda^2 V & \rTo^{H(\tilde d')} & V \oplus V & \rTo^{H(\tilde d)}&\IC^3&& && \cr 
% & && && \uTo_{} && && && \cr 
% & && &&    0    && && && \cr 
% & && && \uTo_{} && && && \cr 
% & && &&    0    && && && \cr 
%& && && \IC^3 & \rTo^{H(d)} & V \oplus V & \rTo^{H(d')} &\Lambda^2 V \,
%\enddiagram 
}
with zeroes and arrows going up and right to fill; $H(maps)$ 
is to remind that these
arrows are induced on cohomologies by $maps$. The 
 $(0,0)$ harmonic forms with values in $B_1^* \otimes B_3$, or by a 
little abuse of language their Serre duals in the upper left corner,
$(0,3)$ harmonic forms with values in $B_3^* \otimes B_1$, are nothing but
what we called earlier $y_{ij}$; the 
$(0,0)$ harmonic forms with values in 
$(B_1^*\otimes B_2) \,\oplus\, (B_2^*\otimes B_3)$ (or again their duals) are
what we called collectively earlier $X^i$, and that we now call more precisely
$(x^i, x'^i)$. We use lower-case letters here to distinguish these deformations
from the initial-point value of the fields, as we shall see shortly.

Second approximation is now to compute the cohomology of the horizontal maps
in \dcf. 0-th line is more explicitly
\eqn\zeroth{
\eqalign{
\IC^3 \ \ \mapr &\qquad \qquad V\oplus V \qquad\qquad
\mapr \ \ \Lambda^2 V \cr
(\alpha,\beta,\gamma)\ \ \mapsto &\ \ 
(\alpha - \beta) X_0, (\beta - \gamma) X_0')\cr
&\qquad\qquad\qquad
(v,v')\qquad\quad \mapsto \ (X_0' \wedge v + v' \wedge X_0)\ ,} 
}
where with hindsight we call $X_0,X_0'$ the global sections of 
$\Hom(B_1, B_2), \Hom(B_2, B_3)$ 
which represent the maps in the original complex $C$. It is easy to see that
cohomology of \zeroth\ 
is $(1,4,6)$. Third line is but the dual complex to this,
and as a consequence its cohomology is $(6,4,1)$.
We can now display our second approximation to the cohomology of the double
complex as 
\eqn\dcs{E^2_{p,q} \equiv 
\matrix{\xymatrix{\boxeqn{6} \ar[r] \ar@{-->}[drr]^{\delta_2}
\ar@{-->}[ddrrr]_{\delta_3}\ar@{-->}[dddrrrr]_{\delta_4}
& 4 \ar[r] & 1\\
&\boxeqn{0}& 0 \ar[u] \\ 
&& \boxeqn{0} \ar[u] &\\
&& 1 \ar[u]\ar[r]  & \boxeqn{4} \ar[r] & 6}}
%\diagram[Postscript=dvips]
%& 6 & \rTo_{} & 4 & \rTo_{}& 1 && && \cr 
% & && && \uTo_{} && && && \cr 
% & && &&    0    && && && \cr 
% & && && \uTo_{} && && && \cr 
% & && &&    0    && && && \cr 
%& && &&  1 & \rTo_{} & 4 & \rTo_{} & \Lambda^2 V \ .
%\enddiagram 
}
Third approximation is to compute the cohomology of a new sort $\delta_2$
of oblique 
arrows, going down 1 and right 2, as shown in \dcs\ (the other two oblique
arrows and the boxes are there for future reference). 
This is defined as follows. Take for example as 
source space the upper left corner of \dcs. These are the 6 harmonic 
three-forms $y$ with values in $B_3^*\otimes B_1$\foot{Concretely, these can 
be expressed in terms of holomorphic sections $y_{[ij]}$ of $B_1^*\otimes B_3$ 
as $\Omega \, (y, \cdot)$, where $(\,,\,)$ is a hermitian metric for the 
bundle: 
in components, $\Omega y_{[i'j']}h^{[i'j'][ij]}$. The sections $y_{[ij]}$
themselves can be got instead easily looking at the definitions of the 
bundles $B_i$ and at the Euler sequence.} that are in the kernel of the 
horizontal map $H(\tilde d')$ -- in particular they are {\it not} among the 4 
three-forms with values in $B_3^*\otimes B_2 \oplus B_2^*\otimes B_1$ that
are displayed in \dcs; they are zero in this group. Being zero in this
cohomology group means being $\bar\partial$ closed; we can thus find a 
two-form $\alpha$ with values in $B_3^*\otimes B_2 \oplus B_2^*\otimes B_1$, 
such that $\bar\partial \alpha = \tilde d' y$. 
We define $\delta_2 y$ now as the result of applying next horizontal map:
$\delta_2 y \equiv \tilde d \alpha$. Summarizing, this is a map between 
the 6 elements in the upper left corner, call
it position $(-2,3)$, and the elements in position $(0,2)$. As this cohomology
group in $E^2$, \dcs, is zero, this map is trivially zero. Looking at \dcs\ 
a little more one discovers that there actually any other arrow of the type
$\delta_2$ (going right 2 and down 1) will be zero by this trick. So our 
next approximation, which would consist in replacing $E^2_{p,q}$ by its
cohomology under the $\delta_2$ maps, is actually the same as $E^2$ again.

It would be wrong, however, to conclude that the approximation procedure has
finally converged. Indeed, again, the image under $\delta_2$ is zero not as
a form, but as an element in the cohomology $E^2$. This means that we can 
go on with our descent procedure: 
\eqn\summar{
\xymatrix{ y \ar@{-->}[r]& \tilde d'y\ar@{-->}[d]\\
&\alpha \ar@{-->}[r] & \tilde d\alpha\ar@{-->}[d]\\
&&\beta \ar@{-->}[r] & d\beta\ar@{-->}[d]\\
&&&\gamma \ar@{-->}[r] & d'\gamma \ ;}
}
by definition, $\bar\partial \beta = \tilde d \alpha$, and 
$\delta_3 y \equiv d\beta$. Since again this ends
on a zero on $E^3 = E^2$, then we can go on and define 
by $\bar\partial \gamma = d\beta$ the map $\delta_4 y = d'\gamma$. 
Compare now
\dcs: this does not end now on a zero, but again on 6. So it is potentially
non vanishing. The example for $\delta_4$ was cleverly chosen: this is 
the only arrow having a chance to be nonzero. It is moreover relevant to our 
problem: in \dcs\ we have boxed the cohomology groups which can contribute to
moduli (they are in position $(p,q)$ such that $p+q=1$, and we have recalled
earlier deformations are in first total cohomology group). From this point
on, arrows will be always too long and will therefore always be zero: so 
our approximation $E^5_{p,q}$ (which is essentially nothing but modifying 
$E^3$ by computing the cohomology of the map $\delta_4$) will be the final one, 
$E^5= E^\infty$\foot{
In general even this final approximation is not 
exactly what one wants: The direct sum of the groups along the boxed diagonal
in $E^5$ would not be 
exactly the first cohomology group of the original double 
complex \linedc. Working with vector spaces we will however see no difference.
Note that for $B_i$ spherical bundles as in this case, things can be anyway 
made more precise building an exact sequence, similar to the Gysin sequence
for sphere bundles, summarizing the content of the spectral sequence.}.
At the end of the day, the number of our moduli is $4+{\rm ker}(\delta_4)$. 
Although 
we have not performed the convoluted computation, the lack of reasons to 
its vanishing at first steps and symmetry reason makes one suspect it is
actually an isomorphism.

The map $\delta_4$ was evidently what was not taken in account in the preliminary
analysis in subsection \subsechyp. To solve the apparent contradiction we had
there, exemplified by the example \equival, clearly it is $\delta_4$ that 
has to lift the extra fields, all the remaining $Y$ in that case. This should
also agree with the other argument we gave, the presence of a further term
in the superpotential.
To look at this in more detail, let us reexpress our computation 
in a way more suitable to generalizations. 

At the stage \dcf, we were faced with harmonic forms
$x_i,x'_i, y_{[ij]}$, following the notation introduced there. 
We can as well consider them as elements $x,x',y$ of $V \oplus V \oplus 
\Lambda^2 V$ and 
express more significantly our next formulas in terms of usual exterior 
algebra, with wedges and contractions. Going to $E^2$ meant
to compute the cohomology of the horizontal arrows. For $x, x'$ this 
only means to consider the ones which satisfy (looking back at first map in
\zeroth) $x'\wedge X_0 + X_0' \wedge x=0$, modulo 
$(x,x')\cong (x,x') + (X_0 A_1 - A_2 X_0, X_0' A_2 - A_3 X_0')$;
in the example considered so far the $A_i$ are still numbers. As for the
$y$, the condition of being in the cokernel can be equally well expressed
as being in the kernel of the dual map $\tilde d'$: this means 
$X_0 \cdot y = 0 = y \cdot X_0'$ (in components, 
$X_0^i y_{ij} =0 = y_{ij} X_0'^j$).

In the following steps there were no essential changes to this picture,
but for the map $\delta_4$. We can see
from the summarizing scheme \summar\ that it involves four horizontal maps,
and each horizontal map is linear in $X_0,X_0'$. In fact, coming back
to the discussion around \summar, one can reexpress the result as 
$$
\delta_4 y = d'\,\delbar^{-1}\left( 
d\, \delbar^{-1} \left(\tilde d \,\delbar^{-1} \left(\tilde d'\, y
\right)\right)\right) \ ;
$$
since each of the $d,d'\ldots$ is linear in $X_0$ and $X'_0$, we see
that this is 
nothing but a sum of Merkulov diagrams of the sextic form in figure 3! 
We can denote the ensuing relation 
symbolically\foot{In the sense that we have not yet established
the way in which the indices are contracted, and whether the whole thing
is actually non-vanishing: the expression only shows how many of the 
various fields should appear.} as $(y\,X_0^2\, X_0'^2)_c=0$. 
The subscript $_c$ is to distinguish this from
a slightly different object we will meet shortly.

When generalizing the three-term complex analyzed so far to general three-term
complexes with general $n_i$ (not necessarily $n_1=n_2=n_3=1$), first change
in previous discussion is that $x, x', y$ become now also matrices, 
gaining extra
indices ($x_{a_1\, a_2}, x'_{a_2\, a_3}, y_{a_3\,a_1}$, where 
$a_i=1,\ldots,n_i$) so that they are elements now of 
$V\otimes \End(n_1,n_2)$, $V\otimes \End(n_2,n_3)$ and $\Lambda^2 V \otimes
\End(n_3,n_1)$ respectively. 
But apart from this, previous expressions were written
in such a way that hold true in this more general case, with the only proviso
that the products are now tensor products of (wedge and contractions in the
exterior algebra of $V$) $\otimes$ (matrix multiplication).
Summarizing the conditions, we have that our deformations are described by
\eqn\geomcond{
\eqalign{
(x,x',y)\quad \rm{such\ that}& \quad 
x'\wedge X_0 + X_0' \wedge x =0\ , \quad X_0 \cdot y = 0 = y \cdot X_0' \quad 
(y\,X_0^2\,X_0'^2)_c =0\cr 
\rm{and\ modulo}&\quad
(x,x',y)\sim (x,x',y) + (X_0 A_1 - A_2 X_0, X_0' A_2 - A_3 X_0',0) \ .}
}

We will see now that the same deformations, {\it but for the quintic 
relation in} \geomcond, come from the quiver in figure 1
with cubic superpotential $\tr (Y_{[ij]}X_iX_j)$.  Indeed, this gives us 
relations
$X'\wedge X=0=X\cdot Y= Y\cdot X'$. Moreover, different
representations are also considered to be equal if they go into each other
by reparameterizations of the three vector spaces $\IC^{n_i}$: 
infinitesimally this yields $(X, X', Y)\sim (X_0,X'_0,Y_0) + 
(X_0 A_1 - A_2 X_0, X'_0 A_2 - A_3 X'_0, Y_0 A_1 - A_2 Y_0) $.
To analyze deformations, we use an apparently ``naive'' procedure which is
actually nothing but a translation of the so-called standard complex (for
categories) \mitchell; we sketch an account of it in appendix a.  
That is, we simply write the fields in the form of 
a ``background value'' plus infinitesimal deformations:
$(X,X',Y)=(X_0 + x, X_0'+ x',y)$; of course this requires that $X_0' \wedge X_0=0$. 
Inserting this into our relations, we find 
$$
(X_0'\wedge X_0)+ X_0' \wedge x + x' \wedge X_0= 0 = X_0\cdot y = y \cdot X_0' \ ;
$$
first piece is there only for illustration, and as we have said it vanishes.
The identifications become likewise $(x,x',y)\sim(x,x',y)+ 
(X_0 A_1 - A_2 X_0, X_0' A_2 - A_3 X_0,0)$. Putting all together, we have exactly
the same relations we had in \geomcond\ --- but for the quintic relation at
the end of first line. 

We have, however, already seen in subsection \merone\ that a deformation 
of the superpotential {\sl is} there; moreover this deformation seems to 
have exactly the same origin as the map $\delta_4$. It is a little subtler to see
more precisely the agreement.
Adding to the initial cubic superpotential $\tr (Y_{[ij]}X_iX_j)$ a 
sextic term $Y^2 X^2 X'^2$, we can see that for instance first relation becomes
$X'\wedge X + Y X^2 X'^2=0$, which infinitesimally gives
\eqn\finally{
 X_0' \wedge x + x' \wedge X_0+ y \,X_0^2 \,X_0'^2= 0\ .
}

The quintic piece in \finally\ has been denoted differently from the one in 
\geomcond, though we just argued they come from the same diagrams. Moreover,
the two quintic objects appear differently: alone in \geomcond, and summed to
something else in \finally. The reason for both things is as follows. 
The spectral sequence map $\delta_4$ takes values not exactly in the space of all
the $y_{ij}$, but in its quotient by the image of the 
map $H(d')$, as we denoted in \dcf.
So, once we have the brute result of the computation of the Merkulov diagram,
which is what we denoted as $y \,X_0^2 \,X_0'^2$ in \geomcond, imposing that this
is zero means only, strictly speaking, that it is zero modulo the image of 
$H(d')$; that is, that this is zero modulo a term 
$X_0' \wedge x + x' \wedge X_0$, for some $x$ and $x'$. A dual method to
deal with the quotient space is to ``fix the gauge invariance'' and
project to a subspace transverse to the orbits of the image of
$H(d')$. The natural choice is to take exactly the same condition of 
transversality that the $y_{ij}$ satisfy,
$X_0^i y_{ij} = 0 = y_{ij} X_0'^j$; this transversality fixes the
gauge because the complexes in zeroth and third line of \dcs\ are
adjoint to each other, and kernel in one is dual to a cokernel (which
is a quotient) in the other. So one has to subtract something to the
``rough'' $(y\, X_0^2\, X_0'^2)$ in order to make it transverse to $X_0$ and
$X_0'$, $X_0^i \,(y\, X_0^2\, X_0'^2)_{c\,ij}=0=(y\, X_0^2\, X_0'^2)_{c\,ij}X_0'^j$.
In general, this can be accomplished fixing metrics 
$\langle\, ,\,\rangle$ on the spaces of the vector spaces of the
$X$,$X'$ and $y$ and writing
\eqn\project{
\delta_4(y)=(y\, X_0^2\, X_0'^2)_c= 
(y\, X_0^2\, X_0'^2) - \langle X'\wedge ((y\, X_0^2\, X_0'^2)\cdot X'), \cdot \rangle
 - \langle (X \cdot (y\, X_0^2\, X_0'^2))\wedge X, \cdot \rangle\ .
}
This formula will become complete when we will write down what $(y\,
X_0^2\, X_0'^2)_{ij}$ is in next subsection; we will look then at examples
to clarify this discussion.

As to the $x,x'$, they are trivially never lifted in neither approach 
(once the relation  $X_0' \wedge x + x' \wedge X_0=0$ is satisfied) : this is 
because such moduli are always trivially solutions of \finally. Let us also
note that, if we had been computing deformations of the same object but on 
$\P^4$, before restricting to $\CM$, we would have had only zero-th line in
\dcs; so the spectral sequence machinery would have stopped at that stage,
and the only solutions we would have found would have been exactly these ones
due to the $X$. So the $X$ are the moduli that were already present in $\P^4$,
and the $Y$ are candidate new moduli. This reflects the splitting between
these two kinds of moduli that we saw in the exact sequence \enhbasseq. 
In fact, although we will not show this in detail here, one can see that the
spectral sequence of this subsection leads exactly to the \enhbasseq. This is
because, for spectral sequences for which only two lines are nonvanishing,
one can summarize the whole information of the spectral sequence in a more
ordinary-looking exact sequence. This happens for instance when computing
cohomology groups of sphere bundles; in that case the resulting exact sequence
is called after Gysin \refs{\botttu}. In this case, exactly the same procedure
would lead to \enhbasseq\ (note again that we called indeed our basic bundles
$B_i$ spherical!); for $D=3$ there is indeed a map between 
$\Ext^2_{\P^4}(E,E)$ and itself, which is nothing but our friend $\delta_4$. 
The reason for which we chose to go through the spectral sequence anyway
is to elucidate the origin of $X$ and $Y$, as we have seen, and to render
transparent the Merkulov origin of $\delta_4$.

Summing up, 
we have two geometrical ways giving the same result: one coming from
a more classical spectral sequence argument, another coming from a more modern
perspective of $A_\infty$ structures. 

\subsec{Obstructions}
We have so far discussed infinitesimal deformations. These
are already connected to higher order terms in the superpotential, since
we are expanding around a complex with non-zero maps. The power 
of the superpotential is indeed that deformations of a vacuum are connected
to obstructions of another one. 

Here we would like to push this a little further by exploring obstructions
to the deformations of the complexes considered so far. It is a usual
phenomenon that infinitesimal deformations do not always integrate to 
finite ones; from the point of view of the moduli space this is simply
signaling a singularity, as the dimension of the tangent space does not
equal the dimension of the manifold. In the case of a bundle, a standard 
reasoning says that deformations are $(0,1)$ forms $a$ 
that obey $\bar\partial a=0 $, where as usual $\bar\partial$ includes the 
starting holomorphic covariant derivative on the bundle under consideration;
and that first obstructions are $a \wedge a =0$ in cohomology. (Higher 
obstructions are called Massey products and can be again found implicitly in 
the $A_\infty$ expression for the superpotential.) Since a bundle can be
also considered as a complex in derived category, this can be rephrased more
generally saying that a deformation $a \in \Ext^1(C,C)$ has as first
obstruction its image under the Yoneda pairing
$$
\Ext^1(C,C)\times \Ext^1(C,C)\mapr \Ext^2(C,C)\ .
$$
In terms of the representation for the $\Ext^i$ that we used in section
\thedouble\ (cohomology of forms with values in complexes), this map
can be expressed in the following way. If $(a_1,a_2,a_3,a_4)$ is an element of
the double complex (where 
$a_i\in\Omega^{0,i-1}(\CM,\oplus B_i^*\otimes B_{2-i})$), its closure means
that $d a_1 =0$, $\delbar a_1 = d a_2$, $\delbar a_2= d a_3$, 
$\delbar a_3 = d a_4$, indicating collectively all our horizontal maps with 
$d$. Then one can see that the correct cocycle representing
the image in $\Ext^2(C,C)$ is, apart from signs,
 $(a^2, ab+ ba, b^2+ac + ca, ad + da + bc +cb)$.
Now, since we know the form of the cocycles associated to our modes 
$y$ and $x$, we can use this to derive higher orders in the superpotential.
For instance, the cocycle associated to $y$ is
$(y, (G\bar\partial^\dagger) d y, 
(G\bar\partial^\dagger) d (G\bar\partial^\dagger) d y, 
(G\bar\partial^\dagger)d(G\bar\partial^\dagger)d(G\bar\partial^\dagger) d y)$.
Using the expression for the ``square'' cocycle we just gave will give higher
and more complicated diagrams {\`a} la Merkulov.

\newsec{Gepner model and linear sigma model}

We now compute the sixth order term in the superpotential using
physics methods.  We follow the general approach of Diaconescu and
Douglas \dd, who rederived and generalized results of Recknagel and
Schomerus \rs\ on boundary states in Gepner models using a simplified
linear sigma model approach.  Again, our concrete considerations are
only for the quintic, but the method is general.

We start with the LG phase of the linear sigma model, which realizes
the orbifold $\BC^5/\BZ_5$.  There is also a world-sheet
superpotential, but as found in \dd\ and as we discussed, the basic
structure of the boundary states and their world-volume theory can be
understood even without this.  This suggests that the case of non-zero
world-sheet superpotential can be treated perturbatively, as has also
been found by Hori and Vafa in a slightly different context \horivafa.

More precisely, we know from world-volume gauge invariance 
(at least if we do not have all five
fractional branes present) that any corrections to the superpotential
must contain the $Y$ fields, so all corrections vanish
if the $Y$ fields are zero.  In other words, the sigma model
with boundary conditions defined by restricting boundary conditions
from the ambient space is exactly solvable.  So, we are effectively
expanding in powers of the deformation away from this solvable case.

Thus we start with the free $\BC^5$ theory, and the D$0$-brane.  The
massless Ramond sector of this open string theory is found by
quantizing the world-sheet fermion zero modes and applying the GSO
projection; this leads to states in correspondence with the odd rank
$p$-forms in $\BC^5$.  The topologically twisted open string theory
has a three-point function which is obtained by multiplying forms and
taking the coefficient of the top form.

One can apply standard orbifold techniques to this theory to derive a
theory with a boundary state for each irreducible representation of
$\Gamma$ (the familiar ``fractional branes''), a spectrum of massless
fermions obtained by projection, and a three point function obtained
by restriction.

After this preliminary discussion, we are prepared to discuss the
computation of higher order terms in the superpotential from the
linear sigma model.  We will generally follow conventions of
topological open string theory.  In any case we only work with disk
world-sheet, where the differences between this and the physical open
string theory are minor.

The original linear sigma model action can be found in \wittenphases.
The general worldsheet description of D-branes in the LG model
has been discussed in \refs{\govjs,\hiv}\ and for the 
linear sigma model in \refs{\gjs,\hiv,\gj,\hellmc,\hellmck}.

Since the superpotential is independent of K{\"a}hler moduli, we start by
taking the limit $\zeta\rightarrow -\infty$, which takes us to the
LG orbifold point.  In this limit, the field $P$ gains infinite mass
and we can drop it.  We effectively have an LG model on $\BC^5/\BZ_5$,
with action
$$
S = \int d^4\theta \sum_{i=1}^5 \bar{\Phi}^i \Phi^i 
+ \int d^2\theta W(\Phi) + \int d^2\bar\theta W^*(\bar{\Phi})
$$
with superpotential
$$
W = c_{i_1i_2i_3i_4i_5}
 \Phi^{i_1} \Phi^{i_2} \Phi^{i_3} \Phi^{i_4} \Phi^{i_5} .
$$
We always use bar to denote space-time complex conjugation.
The chiral superfields $\Phi^i$ have the standard expansion
\wittenphases
$$
\Phi^i = \phi^i +\sqrt2 \theta^+ \psi_+^i +\sqrt2 \theta^- \psi_-^i 
+ \theta^\alpha\theta_\alpha F + \cdots 
$$

We can slightly generalize the model to take independent coefficients $c$ and
$\bar c$ of $W(\Phi)$ and $W^*(\bar{\Phi})$ respectively.
General considerations of topological theory will tell us that correlation
functions of chiral operators $O(\Phi)$ will only depend on $c$, so
we will set $\bar{c}=0$.

We represent the disc as the upper half plane $z=x+iy$ with
$y\ge 0$.
The boundary conditions of interest to us are Dirichlet boundary
conditions\govjs, which we take to be $\phi=\bar \phi=0$ and 
$\psi_+=\psi_-$, $\bar\psi_+=\bar\psi_-$.  It is useful to form
the combinations 
$$\eqalign{
\xi^i=\psi_+^i + \psi_-^i ;\qquad
\bar\xi^i=\bar\psi_+^i + \bar\psi_-^i \cr
\tau^i=\psi_+^i - \psi_-^i ;\qquad
\bar\tau^i=\bar\psi_+^i - \bar\psi_-^i \cr
}$$
in terms of which the fermion boundary conditions are $\tau=\bar\tau=0$.
Each $\bar{\xi}$ fermion will then have a zero mode (the $\xi$ zero modes
do not survive the topological twisting).

The linear sigma model has  two $U(1)$ R-symmetries of which one linear
combination is preserved by the boundary conditions. Under this 
symmetry,
$\theta$ has charge $+1$, $d\theta$ and $D$ have charge $-1$,
and $\Phi$ has charge $2/5$.
Nonzero correlation functions must include enough fermions to saturate
the $\bar{\xi}$ fermion zero modes; this leads to a background $U(1)_R$
charge $-3$.

The boundary operators we will use are(see Appendix B for
more details)
$$%\eqalign{
O_{(k;0)}^{i_1\ldots i_k} 
%&= \bar{D}\bar{\Phi}^{i_1} \ldots
%\bar{D}\bar{\Phi}^{i_k}|_{\bar{\theta}=0} \cr
= \bar{\xi}^{i_1} \ldots \bar{\xi}^{i_k}
%}
$$
and correspond to the constant $k$-forms $d\bar{\phi}^{i_1}\ldots
d\bar{\phi}^{i_k}$
on $\BC^5$.  They have $U(1)_R$ charge $3k/5$.

The subscript $0$ signifies that this is the lowest
component of a superfield; one also has the one-form
$d\theta O_{k} = O_{(k;1)}$, and a two-form $d^2\theta O$ from the
world-sheet superpotential.  

We denote $O_{(1;n)}=O_{X;n}$ and $O_{(3;n)}=O_{Y;n}$ as in
our previous notations.  Finally, we include the five-form
$O_{(5;n)}=O_{Z;n}$.

The quotient by $\BZ_5$ is handled as in \dm\ and we will generally
suppress these details of the computation from now on.  One should
also think of each boundary operator as coming with Chan-Paton factors,
which can be chosen to specify any desired ordering of operators
along the boundary.  Equivalently, the open string couplings are
free noncommuting variables.  Following these considerations,
each world-volume superpotential term we write is to be interpreted
as a sum over terms, one for each ordering of links which leads to a
closed path on the graph.  

The propagators we will need are then:
$$\eqalign{
\vev{\phi^i(z_1) \p_{y_2} \bar{\phi}^j(x_2)} &=
         -\delta^{ij} \left( {1\over z_1-x_2} - {1\over \bz_1-x_2}\right) \cr
\vev{\tau^i(z_1) \bar\xi^j(x_2)} &=
  \delta^{ij}\left( {1\over z_1-x_2} - {1\over \bz_1-x_2}\right) \cr
}$$
Only derivatives of $z$ appear, so the integrand is single-valued.

The contribution to an
$n$-point correlator at $m$'th order in the superpotential is an
integrated correlation function, with the action of $SL(2,\BR)$
divided out.  The simplest example is the 
leading term in the world-volume superpotential, which comes
from the three-point function
$$
\vev{O_{X;0}|_0 O_{X;0}|_1 O_{Y;0}|_\infty}
$$
producing
$$
\CW = \Tr X^{i_1} X^{i_2} Y^{i_3i_4i_5} \epsilon_{i_1i_2i_3i_4i_5} + \ldots 
$$
as found in \refs{\dd,\bs}. 

The first correction will come from one insertion of the world-sheet
superpotential.  
We will fix the $SL(2,\BR)$ symmetry by 
fixing the positions of two boundary operators to be $z=0$ and $z=\infty$,
and taking the one-form version of the bulk operator, integrated along
a contour $x\in\BR$, $y$ fixed, which contains one point in each
$SL(2,\BR)$ orbit.  This leads to the insertion
$$
O_W = \int_{-\infty+iy}^{+\infty+iy} dx O_{(W;1)}
\int_{-\infty+iy}^{+\infty+iy} dx\ 
c_{i_1i_2i_3i_4i_5} \phi^{i_1}\phi^{i_2}\phi^{i_3} \phi^{i_4} \tau^{i_5}.
$$

Since this is free theory, it is easy to see that this leads to the
following nonzero correlation functions.  The first is
$$
I_1 = \vev{O_{Y;0}^{j_1j_2j_3}(\infty) 
 \int dx_1  O_{X;1}^{k_1}(x_1)
 \int dx_2  O_{X;1}^{k_2}(x_2)
            O_{Y;0}^{j_1j_2j_3}(0)
 \int dx_3  O_{X;1}^{k_3}(x_3)
 \int dx_4  O_{X;1}^{k_4}(x_4)
   O_W}
$$
which corresponds to the term \tenthorder,
and
$$
I_2 = \vev{O_{Z;0}^{j_1j_2j_3j_4j_5}(\infty) 
 \int dx_1  O_{X;1}^{k_1}(x_1)
 \int dx_2  O_{X;1}^{k_2}(x_2)
            O_{X;0}^{k_3}(0)
 \int dx_3  O_{X;1}^{k_4}(x_3)
 \int dx_4  O_{X;1}^{k_5}(x_4)
 O_W }
$$
whose significance we will mention later.  In both cases, the
limits of integration are chosen to preserve the operator ordering.

For $I_1$, for example, the integrand is
%\centerline{\epsfxsize=5in \epsfbox{uhp.eps}}
\eqn\eione{
I_1\propto\int_{-\infty}^\infty dx \int_{-\infty}^0 dx_2 
\int_{-\infty}^{x_2} dx_1 \int_0^\infty dx_4 \int_0^{x_4} dx_3
\prod_{i=1}^4 f(x,y;x_i) f(x,y;0)
}
where $$f(x,y;a)\equiv {{2y}\over{(x-a)^2 + y^2}}$$ is the Lorentzian which
comes from the boundary-bulk two-point functions -- $z=x+iy$ being
the location of the bulk operator and $a$ the location of the boundary
operator. 
There are five such terms in all -- four from the $X$ insertions
on the boundary and one from the $Y$ insertion at the origin. $SL(2,\BR)$
invariance implies that the above amplitude be independent of
$y$. This is equivalent to the scale invariance of the amplitude. The
Lorentzian scales as: $f(\lambda x, \lambda y; \lambda a) = \lambda^{-1}
f(x,y;a)$. This gives a $\lambda^{-5}$ which gets cancelled from the
$\lambda^5$ which comes from the integration measure. The scaling
does not modify the limits of integration as well.

From the expansion of the action, we have to include all allowed
orderings of the $X$'s -- this will lead to a combinatoric factor
which is included below. By allowing the combination  $x_3>x_4$
and a similar thing for $x_1,x_2$ and one can rewrite  the nested
integrals in unnested form (This is not quite necessary --
one can also do the nested integrals to get the same result)
\eqn\eionea{
I_1={1\over4}\int_{-\infty}^\infty dx \int_{-\infty}^0 dx_2 
\int_{-\infty}^{0} dx_1 \int_0^\infty dx_4 \int_0^{\infty} dx_3
\prod_{i=1}^4 f(x,y;x_i) f(x,y;0)
}
where the ${1\over4}$ is what remains of the $1/4!$ after we take various
orderings into account. 

The following change of variable simplifies things: let
$x=y \tan \theta$ and $x_i-x=y \tan \theta_i$. In terms of the
new variables $f(x,y;x_i)=y^{-1}\cos^{2}\theta_i$  and
the integrations over $x_i$ are easily done to obtain
\eqn\eioneb{\eqalign{
I_1 &= {{(2)^5}\over{4}}
\int_{-\pi/2}^{\pi/2} d\theta \left({\pi\over2} -\theta\right)^2 
\left({\pi\over2} +\theta\right)^2 \cr
&= {4\over5} \pi^5 
}}
The contribution of this term to the superpotential takes the form
(suppressing Chan-Paton labels)
\eqn\Enewterm{
I_1\ c_{i_1i_2i_3i_4i_5} \epsilon^{j_1j_2j_3j_4i_5} X^{i_1}X^{i_2}Y_{j_1j_2}
X^{i_3}X^{i_4}Y_{j_3j_4}
}
which is precisely the form required to resolve the contradiction 
mentioned in sec. 2.6.

This expansion can of course be developed to arbitrary order.
It is rather simpler than the usual interacting field theory in
that there are only bulk-boundary correlators.  The generalization
of the term we computed to all orders takes the form
\eqn\superexpansion{
\CW = \epsilon_{i_1i_2i_3i_4i_5} \sum_{m\ge 0} c^m \Tr (XXY)^{m+1}
}
where each further term in the sum over $m$ of course involves a sum
over index contractions, with coefficients determined by doing
integrals of the type we discussed.  Of course,
the resulting coefficients need not be simple.

The $O(X^5)$ term should also play a role, for example in the case
$(1,1,1,1,1)$. The contribution to the superpotential is of the form
(suppressing Chan-Paton labels and the associated trace)
\eqn\eitwoa{
I_2\ Z\ c_{i_1i_2i_3i_4i_5}  X^{i_1}X^{i_2} X^{i_3}X^{i_4}X^{i_5}\sim 
I_2\ Z \ W(X)
}
where $I_2$ is an  integral (non-vanishing) similar to \eionea. 

This term is interesting as we can now see a role for the field $Z$,
analogous to that of the $P$-field in the world-sheet linear sigma
model associated with the same CY.
Namely, the equation of motion $\p\CW/\p Z=0$ enforces the hypersurface
condition $W=0$.
It is tempting to identify this as the term in the potential which
lifts degrees of freedom not lying on the hypersurface.

\newsec{Further examples}

We examine here the other examples we introduced in subsection \subsecex, 
and more. For simplicity we will keep in mind the Fermat case.

We start with the RS state, which corresponds to a complex with 
$\{n_i\}=(0,0,1,2,1)$. The first thing is to write the complex analogous
to \zeroth, which is in this case 
$$ 
\IC^{1 + 2^2 +1} \mapr \IC^5 \otimes \IC^{1\times2 + 2 \times1} \mapr 
\Lambda^2 \IC^5 \ .
$$
We do not {\it a priori} know which maps correspond to the RS state,
but a plausible guess is that $X^i$ is equal to the multiplication with
the vector $e_i\in \IC^2$, and $X'i$ as the contraction by the same vector.  
In this case, the cohomology is $(1,8,3)$. This tells us that on $\P^4$
the complex would have moduli space of dimension 8; we now have to check 
how many of the candidate extra 3 moduli add to these when restricting to 
$\CM$. 

First of all, the equation $X \cdot y = 0 = y \cdot X'$ becomes
now $\epsilon_{ijkl1} y^{ijk}=0= \epsilon_{ijkl2} y^{ijk}$; this leaves only
$y_{12i}$, $i\ne 1,2$: these are the three candidates. The real check is now
again $\delta_4$. In this case the situation is more interesting: in components 
one can see that the quintic term is now 
$(X^2 X'^2 y)^{ij}= y^{1ij} +y^{2ij}$. So the meaningful equations are
obtained setting one of the $i,j$ equal to 1 or 2:
$$
0=x^i_1 + y^{1i2}=x^i_2 + y^{2i1}= x'^i_1 + y^{i12}=x'^i_2 + y^{i21}
$$
(remember that $x$ and $x'$ are each element of $\IC^5\otimes \IC^2$,
and so they have an extra lower index taking values 1 or 2). None of
these equations set a $y$ field to zero: in each case, there is the
possibility of switching on a companion $x$ field in such a way that
the equation is satisfied without having to put the $y$ to zero. We
can again rephrase this in a dual picture: The result $y^{1ij}
+y^{2ij}= [\delta_4(y)]_{ij}$ is not ``gauge fixed'', in the sense
that it is not orthogonal to $X$ and $X'$: those which are are of the
form $y^{12i}$, as we already know. So when we try to project $y^{1ij}
+y^{2ij}$ to the space of the $y^{12i}$, as we did in \project, we get
zero: this is exactly the dual of the fact that these equations can be
``gauged away'' by introducing the extra $x$ and $x'$. Either way,
this means that the 3 fields are in the kernel of $\delta_4$. This
predicts that this complex should have in total $8 + 3=11$ moduli on
$\CM$, which is consistent with the Recknagel-Schomerus computation.

We now come to the other class of examples we introduced in subsection 
\subsecex, that is $\CO(i)_{|_{\P^j}}$. This class of examples is easy
to compute using the Beilinson procedure because the cohomology groups of the
$\CO(i)$ are very well known. In general, it is better to reduce the result
of this procedure to one line, in order to have a simple complex of $B_i$,
which leads to a quiver representation. 
In the class of examples at hand, this happens if $i+j \ge 4$. It turns out
interesting however to also break this rule; the resulting example is 
interesting and we will describe it in a little more detail, whereas
for the other cases we only describe the results.

The first case we handle is $\CO(2)_{|_{\P^3}}$. In this case 
$\{n_i \}=(0,0,1,4,10)$ and so the starting complex 
%(the analogue of \zeroth) 
is $(1+4^2+10^2) \mapr 5\times(1\times 4+4\times 10)\mapr
10\times(1\times10)$; as opposed to the RS case above, we can write
down the maps explicitly here with no guesswork, and the cohomologies
result to be $(1,4,0)$. This means that there are 4 $X$ and no $Y$; 
so it means that to the 4 moduli that the sheaf had on $\P^4$ 
(which are
right, because a hyperplane $\P^3 \subset \P^4$ has 4 moduli and the
$\CO(2)$ on it is rigid) nothing adds when restricting to the
Calabi-Yau. So in this case we did not really need our methods.

Something more interesting arises if one considers instead
$\CO(2)_{|_{\P^2}}$, whose numbers are $\{ n_i\}=(0,0,1,3,6)$. The
complex is $(1+3^2+6^2) \mapr 5\times(1\times3+3\times6)\mapr
10\times(1\times6)$ and its cohomology is $(1,6,6)$. This case is more 
interesting as in addition to the 6 moduli on $\P^4$ 
(again these are right, as
this is the dimension of a Grassmannian $G(3,5)$), there are 6 more
candidate moduli upon restricting to $\CM$. As in the RS case we can
write down the explicit equations and we discover in a similar way
that all the moduli are not lifted. So we have 6+6=12 moduli for this
sheaf on $\CM$. How can we check that this is right geometrically?
Viewing this sheaf as a bundle on its support, as usual we have to
count deformations of the support inside the Calabi-Yau $\CM$ and
deformations of the line bundle on the curve $\P^2 \cap \CM$, which is
a Riemann surface $C$. First type of deformations are at least six, 
as moduli that
were there on $\P^4$ cannot be lifted. On the other hand, $C$ can be
viewed as a curve of degree 5 inside $\P^3$, and so has genus
6. Deformations of line bundles on a Riemann surface are exactly as
many as the degree, and so we have 6 other deformations from here. We
know that 12 is the maximum allowed, and so we get that it is the
right number.

It would now be natural to consider something which in $\P^4$ has
support $\P^1$, in such a way that we get 5 D0 branes on the quintic. 
According to the rule we expounded above, to have a single quiver
representation we should now consider at least
$\CO(3)_{|_{\P^1}}$. This would give a longer complex, examples of 
which we will consider shortly.
We can however
remain in the more familiar domain of complexes of length 3 by 
considering, instead of a single quiver
representation, a more generic complex of quiver representations:
explicitly $q_1\equiv(0,0,1,2,3) \mapr q_2\equiv(1,0,0,0,0)$. 
This is a little off the
track in that an analysis with a double complex similar to what we did
in section \thedouble\ for sheaves shows that this object is not simple
($\Hom > 1$), and deformations may also come from a $\Ext^2(q_2,q_1)$
that we should compute. It is interesting however to check what are
the deformations of the $\Ext^1(q_1,q_1)$. There are in this case 6 $X$
and 9 $Y$; none of these $Y$ is lifted by our superpotential, so that 
the total moduli are 6+9=15. This is perfectly consistent with fact
that the total object represents 5 D0 on the Calabi-Yau.

\subsec{Longer complexes}

So far we have restricted ourselves to examples of complexes for
which only 3 consecutive nodes were present. The reason this is
simpler can be seen in terms of the double complex. For longer
complexes, the complex \linedcf\ would again be present only for $p=0$
and 3, but it will be longer in both directions, because
there will now be bundles $B_5^*\otimes B_1$
and $B_1^*\otimes B_5$. In the general case $(n_1,\ldots,n_5)$, the
zeroth line \zeroth\ will read
\eqn\zerothgen{
\IC^{\Sigma n_i^2}\mapr V \otimes \IC^{\Sigma n_i n_{i+1}}
\mapr \Lambda^2 V \otimes \IC^{\Sigma n_i n_{i+2}}\mapr\ldots
\mapr \Lambda^4 V \otimes \IC^{n_1 n_5} \ . 
}

We will start by looking at the example $(0,1,1,1,1)$, which should
be the anti brane to $B_1$ and thus
rigid on $\CM$. In this case $E^2$ is (compare \dcs)

$$
\matrix{
\xymatrix{4 \ar[r] \ar@{-->}[dddrrrr]_{\delta_4'}&
\boxeqn{6} \ar[r] \ar@{-->}[dddrrrr]_{\delta_4}
& 4 \ar[r] & 1\\
&&\boxeqn{0}& 0 \ar[u] \\ 
&&& \boxeqn{0} \ar[u] &\\
&&& 1 \ar[u]\ar[r]  & \boxeqn{4} \ar[r] & 6 \ar[r]& 4 }}\ .
$$

What happens here is that both $\delta_4, \delta_4'$ are isomorphisms,
and lift both $y$ and $x$ deformations. So we already see from this
example that the comparison to the procedure for computing
deformations and $\Hom$ of quiver representations 
we performed in section \thedouble\ is less
evident. This is not a surprise if we get back at our exact sequence 
\enhbasseq. Remember that, as we outlined at the end of section
\thedouble, we can recover this exact sequence from the spectral
sequence via an argument {\`a} la Gysin for our spherical bundles. Then
we see that a larger spectral sequence gives rise to non-zero 
$\Ext^1_{\IP^4}(E,E')$ that complicate the story exactly as it happens
in lower dimensions, see remarks after \enhbasseq.

We end with two natural continuations of the examples discussed so
far. One is $(1,1,1,1,1)$, for which zeroth line is this time
$1\mapr 4 \mapr 6 \mapr 4 \mapr 1$ and there are three maps $\delta_4$
this time. All of them are isomorphisms, and this time even $\Hom$
gets cancelled! This is not surprising if we remember that this state
is the D0 on $\P^4$, and so it restricts to nothing on $\CM$. 

Our last example will be
the more straightforward realization of the state with 5
D0 branes on $\CM$ that we anticipated above, obtained as $\CO(3)_{|_{\P^1}}$.
This time we get
$$
\matrix{
\xymatrix{4 \ar[r] \ar@{-->}[dddrrrr]_{\delta_4'}&
\boxeqn{9} \ar[r] \ar@{-->}[dddrrrr]_{\delta_4}
& 6 \ar[r] & 1\\
&&\boxeqn{0}& 0 \ar[u] \\ 
&&& \boxeqn{0} \ar[u] &\\
&&& 1 \ar[u]\ar[r]  & \boxeqn{6} \ar[r] & 9 \ar[r]& 4 }}\ .
$$
We have this time that both $\delta_4, \delta_4'$ are zero; this
implies that deformations are $6+9=15$ as in the previous 5 D0
realization; moreover, here we have that the $\Hom$ are
$1+4=5$, so the 5 D0 are not bound, but each contributes its own
unbroken gauge groups.

Let us note that in all the examples we have seen, either all or none
of the $Y$ are lifted. The reason for this is that the
original double sequence has a symmetry $(p,q) \mapr (-p,3-q)$, which
one can use to construct the same arrow in
reverse. This means that as long as this symmetry is not broken the
map (once projected as in \project) can be written as
$[\delta_4(y)]_{ij}=\lambda\, y_{ij}$, so is either the zero map or an 
isomorphism.

\subsec{Conclusions}

We saw in these examples that the framework as we described it appears
to be a satisfactory description of bound states involving three types of
fractional brane.  It is not obvious at this point whether
higher order corrections of the form \superexpansion\ are required to get
exact results.  Besides a direct approach, one might gain information
on this point by comparing our results with those of \dgr, who get
information on the spectrum of BPS branes from a very different
starting point (supergravity and the attractor mechanism).

We discussed bound states of more fractional branes as well.
Mathematically, these
examples work rather similarly to the previous ones, with the non-trivial
aspects of the restriction captured in the maps $\delta_4$, information
which we believe can again be summarized in the superpotential.
What changes is only that the pairing now involves
additional $\Hom$'s between the complexes, which appear on restriction.

In fact, a naive approach to modeling these bound states
as supersymmetric quiver gauge theory involving four or five nodes, does
not work.  The basic problem is that one does not see the extra $\Hom$'s in
this description.  These are additional brane-antibrane tachyons which
must appear because of identifications such as $B_1+B_2+B_3+B_4=\bar B_5$.
Similar problems with the naive application of supersymmetric gauge theory
to problems with too many different fractional branes were already observed
in \dd; in particular one cannot reproduce the correct bosonic masses
using Fayet-Iliopoulos terms.

One way to deal with this would be to forget about supersymmetric
gauge theory (i.e. abelian categories) and instead always work with
the derived category.  Another possibility is to apply Seiberg
dualities to some of the fractional branes \berdou, to try to get valid
supersymmetric field theories in these cases.  Clearly systematizing
this is an important question for future work.

\newsec{Conclusions and further directions}

In this work, we have taken a further step in the development of a
concrete framework which can describe all BPS D-branes in Calabi-Yau
compactification of string theory, by showing how to handle a compact
Calabi-Yau.

The framework can be motivated and justified both from physics and
mathematics starting points, and we have shown at length how the two
points of view are related.  Not being mathematicians, we will
summarize what we have done from this point of view very succinctly.
In principle, the
obstruction theory of the restriction of the derived category of coherent
sheaves from $\P^4$ to a quintic hypersurface, could be obtained from
computations in holomorphic Chern-Simons theory.  In practice such computations
are difficult.
In section 4 of this work, we define a topological field theory 
which encodes the same information, but in which computations are
straightforward, and exhibit the first non-trivial correction.

From a physics point of view, we start with certain fractional branes
at the Landau-Ginzburg orbifold point of the linear sigma model.
These are a particularly simple set of constituents from which a large
subset of all branes can be constructed as bound states, by finding
vacua of their supersymmetric world-volume theories (and
generalizations of this prescription).
In previous work, it was shown that by taking a cubic world-volume
superpotential, one precisely describes the branes which correspond to
sheaves on the resolved orbifold, corresponding to the linear sigma
model with zero world-sheet superpotential.  
Adding the world-sheet superpotential constructs the CY as a
hypersurface in this resolved orbifold, and our primary question was to
find out whether the same type of description is valid and what precise
modifications the world-sheet superpotential induces.

We studied the restriction to the hypersurface in some detail using
homological algebra and spectral sequence techniques, justifying the
claim that in three and more complex dimensions branes on CY are very
similar to those on the ambient space, so that the world-sheet
superpotential can be treated perturbatively.  From a physics point of
view this might be considered surprising, as the world-sheet
superpotential induces a non-trivial RG flow from (in our case) $\hat
c=5$ to $\hat c=3$.  Nevertheless it clearly works in the problem at
hand, and this is one of the important lessons of our work.

Perhaps the most conservative explanation of why perturbation theory
in the world-sheet superpotential works here, is that our D-branes are
all localized to the orbifold singularity, and after the resolution
become branes wrapping cycles in $\P^4$; thus the sector of the theory
we consider has fewer light degrees of freedom than a generic $\hat
c=5$ string.  Of course it might be that it works in greater generality.

The restriction to the CY does change the detailed properties of the
branes.  On physical grounds, most of these modifications can be
expressed by corrections to the world-volume superpotential.  We
computed the leading such correction, which contains the leading
dependence on the complex structure of the CY, and showed both on
general grounds and in examples that this correctly describes the
modifications to the category of coherent sheaves produced by
restriction.  Thus, we see that branes on compact CY's can be
described by these methods.

Although we only discussed the case of the quintic, as we mentioned
in the introduction these methods generalize to large classes of CY's
which can be realized in the linear sigma model framework, and
the new linear sigma computations we have done appear no harder in the
general case.

Can we hope that these techniques could lead to a complete description
of branes on CY?  There are numerous issues to be understood.

The first is that strictly speaking one only gets a subset of the
K theory, those classes which can be obtained by restriction.  This
is probably a minor problem as one can take limits in which these objects
become reducible, splitting into objects with the other classes.
For example, although one cannot obtain the D$0$ on the quintic, one
can intersect a curve with the defining surface to obtain five D$0$'s;
the formalism then produces the additional moduli required to move these
independently.

The second problem is that one is still working by expansions around
more solvable points, and some phenomena are hard to see this way.
For example, one expects moduli spaces of branes on compact CY to have
some type of compactness.  In this description, the compactness of
moduli associated to the ambient space (the $X$'s) is manifest;
however one also expects compactness in the $Y$ directions, which
correspond to field identifications which will not be obvious.  At
this point, this is probably the most serious problem we foresee.
Such field identifications will be signaled however by new light
states (e.g. consider the theory of two D$0$'s on a torus) which might
well have some signature in the superpotential, so it is too early to
say whether this can be better understood.

In section 5 we mentioned the problem that quiver theories including
too many different fractional branes cannot be treated in a naive way
as supersymmetric gauge theories (a signal of this is that the sum of
bosonic squared masses around a closed loop is non-zero, which cannot
be reproduced by Fayet-Iliopoulos terms).
We believe that this problem goes away upon more correctly
treating brane-antibrane combinations, either in the derived category or
by performing appropriate Seiberg dualities.  This point also deserves
to be better understood, but is probably not an obstacle.

Finally, of course a complete solution of this problem with these
ideas would seem to rest on being able to find the exact
superpotential.  Now, there are many highly nontrivial exact
superpotentials known for $\CN=1$ supersymmetric theories, so in itself
we consider this a rather encouraging reformulation of the problem.

As we discussed, this superpotential can also be considered as a
rewriting of the holomorphic Chern-Simons action in a more explicit
form, depending on a finite number of variables.  We have essentially
proposed an alternate topological open string theory which is more
tractable than the original one.
This theory is rather similar to that developed by Cattaneo and Felder
\catfel\ to describe deformation quantization, and particularly to their
generalizations with higher rank forms.  Many mathematical 
connections between these problems are known. \seidel

There is also a close analogy between this superpotential and the
periods of the holomorphic three-form which were the original subject
of mirror symmetry.  Indeed, the superpotential for a wrapped two-brane
is a semiperiod \refs{\Donaldson,\wittenqcd}\ which in interesting examples
\agvafa\ satisfies generalized Picard-Fuchs equations for which the periods
are also solutions \refs{\gjsinst,\mayr}.  
Various extensions of this relation are known, 
suggesting that the coefficients of the series expansion
\superexpansion\ should be closely related to those in the comparable
series expansions of the periods
\candelas, or that the superpotential itself will satisfy a linear
differential equation.

For all of these reasons, we believe that the further study of this
superpotential will be a fruitful direction in the near future.
Here we explained how this would lead to 
a concrete description of the derived category of sheaves on CY,
which would have numerous applications in physics and mathematics.

\medskip

We acknowledge valuable discussions with D.-E. Diaconescu, R. Dijkgraaf,
K. Paranjape, and especially with P. Seidel.  We also thank B. Szendroi
for critical remarks, which are addressed in version 2.
T.~J. and A.~T.~would like to thank
the NHETC of Rutgers University for its kind hospitality 
during first phases of this project.

This work was supported by DOE grant DE-FG02-96ER40959.

\appendix{A}{The standard complex for quivers}

We describe this here both to show that our ``intuitive'' way of
computing deformations of quivers does not hide mathematical
subtleties, but also to show how to compute higher $\Ext$. The degree
of complication we will meet should make one appreciate the power of
generalized McKay correspondence, that allows one to compute such
groups by means of more standard geometrical procedures.  We will
start from abstract definitions and try to get quickly to examples, to
show how it works.

Given our quiver $Q$, let us denote its nodes by $d_0,\ldots,d_k$; as for 
arrows between $d_i$ and $d_j$, 
we will consider them as formal generators of abelian groups; 
this group we dub $Q(d_i,d_j)$. (This realizes what is called a 
$\IZ$-linear category.) 
The standard complex is then 
$$
\eqalign{\ldots 
\oplus_{i,j}Q(d_i,d')\otimes Q(d_j,d_i)\otimes Q(d,d_j)&\mapr
\oplus_i Q(d_i,d')\otimes Q(d,d_i)\mapr Q(d,d')\cr
a\otimes b \otimes c \quad &\mapsto \quad ab \otimes c - a \otimes bc}
$$
where all tensor products are over $\IZ$, and a map has been shown for 
illustration. The importance of this complex is that, given any quiver
representation $M$, tensoring with this complex gives a projective resolution
of it:
\eqn\stancompM{
\oplus_{i,j}Q(d_j,?)\otimes Q(d_i,d_j)\otimes M(d_i)
\mapr \oplus_i Q(d_i,?)\otimes M(d_i)\mapr M\ ;
}
here, for instance, central term is a representation whose vector space
corresponding to node $d$ is $\oplus_i Q(d_i,d)\otimes M(d_i)$. We will see
in a moment how this works. Before of that, let us also recall that the 
$\Ext^i$ of two quiver representations $M$ and $N$ are then obtained 
taking $\Hom(\cdot,N)$ of \stancompM, and considering the cohomology of
the resulting complex. We obtain:
\eqn\stancompH{
\eqalign{\Hom&(M,N)\mapr\cr
&\xymatrix{\Hom(\oplus_i Q(d_i,?)\otimes M(d_i),N)\ar[r]^\delta\ar@{=}[d] &
\Hom(\oplus_i Q(d_j,?)\otimes Q(d_i,d_j)\otimes M(d_i),N)\ar@{=}[d] \\
\oplus_i \Hom(M(d_i),N(d_i))\ar[r]& 
\oplus_{i,j}\Hom(Q(d_j,d_i)\otimes M(d_j),N(d_i))\ .
}}
}
and so on.
The two vertical equalities are because of something called Yoneda lemma. 
We will illustrate it in a baby example, which will takes us on our way to 
more grown-up ones shortly. 
Consider a quiver $d_1 \mapr^\alpha d_2$, 
and the representation of it $Q(d_1,?)$, where $d_1$ 
is its first node. This 
representation is simply the one whose vector space at each node is generated
by the arrows starting from $d_1$; in this case, at $d_1$ 
itself we only have $e_{d_1}$, the identity on $d_1$, whereas at
$d_2$ we have $\alpha$. This representation can be thus written more 
concretely as $1 \mapr^1 1$. If we consider now quiver morphisms of this into
any other representation $N$,
$$
\xymatrix{N(d_1) \ar[r]^{N(\alpha)}& N(d_2)\\
1\ \ar[u]\ar[r]^1 &\ \ 1\ar[u]\ ,}
$$
then we have as our only choice the image
of the lower left 1 inside $N(d_1)$, call it $n_1$. Indeed 
the image of the other 1 then follows to be simply $n_2=N(\alpha)(n_1)$ since
the diagram should commute. So we have got an explicit isomorphism
$\Hom(Q(d_1, ?), N) \buildrel\sim\over\mapr N(d_1)$; more generally 
the same mechanism yields
$$
\Hom(Q(d,?),N)\buildrel\sim\over\mapr N(d)
$$
for $Q$ whatever quiver and $d$ whatever node.

What we want to do now is to illustrate the maps in \stancompH, so that
it become explicit to compute its first cohomology. We consider now
the quiver $d_1 \mapr^\alpha d_2 \mapr^\beta d_3$. We start
by studying the map $\delta$ in \stancompH. Its source space is in this case
given by homomorphisms of representations:
\eqn\simpleHom{
\xymatrix{
N_1 \ar[r]^{N(\alpha)}&N_2 \ar[r]^{N(\beta)}&N_3 \\
e_1\otimes M_1 \ar[u]^{h_1}\ar[r]& 
e_2\otimes M_2 \, \oplus \, \alpha\otimes M_1 \ar[u]^{(h_2, N(\alpha)h_1)}
\ar[r]&
e_3\otimes M_3 \, \oplus \, \beta\otimes M_2 \,\oplus\,
\beta\alpha\otimes M_1\ . \ar[u]^{(h_3, N(\beta)h_2,N(\beta\alpha)h_1)}\\}
}
Here the $h_i$ can be simply seen as elements of 
$\Hom(M(d_i),N(d_i))$, as another manifestation of Yoneda lemma, 
compare again with \stancompH. Our aim is then to find the image of this.
Elements of the arrival space of $\delta$ are now multiples 
$(a_1,a_2,a_3,b_1,b_2,c_3)$, which can be seen as candidate deformations (in 
the case $M=N$, of course) of the given representations for $e_1$, $e_2$, 
$e_3$, $\alpha$, $\beta$, $\beta\alpha$. 
The map we are going to find is then the
identification of different deformations; we will consider afterwards their
consistency as candidate deformations (closure under the following map).
Also for these would-be deformations $a_i$, $b_i$, $c$ a representation 
of the type we gave in \simpleHom\ for $(h_1,h_2,h_3)$ can be given. On the 
lower right corner we have now
\eqn\lesssimple{
e_3 \otimes e_3 \otimes M_3 \,\oplus\, e_3 \otimes \beta \otimes M_3
\,\oplus\, \beta \otimes e_3 \otimes M_3 \,\oplus\, 
e_3\otimes \beta\alpha \otimes M_3 \,\oplus\, \beta \otimes \alpha \otimes M_3 
\,\oplus\, \beta\alpha \otimes e_3 \otimes M_3 
}
which goes into $N_3$ through a map $(a_3,b_2,N(\beta)a_2, c, N(\beta)b_1,
N(\beta\alpha) a_1)$ (compare again with \simpleHom). An element 
$e_3\otimes e_3\otimes m_3$ in the first summand of
\lesssimple\ is now sent by the map in the standard complex to 
$(e_3 e_3 \otimes m_3 - e_3 M(e_3) \otimes m_3)=0$. An element 
$e_3\otimes \beta \otimes m_3'$ in second
summand is already more interesting, being sent into $\beta\otimes m_3'\oplus 
(- e_3 M(\beta)m_3')$. To this we can apply third vertical map in \simpleHom\ 
getting $N(\beta) h_2 - h_3 M(\beta)$! Here we recognize nothing but the
infinitesimal action of change of basis on nodes $d_2$ and $d_3$ 
on the representation of the arrow $\beta$ (again, in view of the case $M=N$ 
which is our main application here).
Going on in this way we find as image of $h_i$ the multiple defined by
$$
a_i=0\ , \quad b_1 =N(\alpha) h_1 - h_2 M(\alpha)\, \quad 
b_2=N(\beta) h_2 - h_3 M(\beta)\ ,\quad 
c= N(\beta\alpha) h_1 - h_3 M(\beta\alpha)\ . 
$$

This is good, but the really interesting thing is to check the way deformations
are considered as valid ones, more than the identifications among them.
A cursory look at what we just described will convince anyone that writing
everything explicitly would be not particularly readable. But we can describe
salient pieces. We start now from a multiple $a_i$, $b_i$, $c$, and look for
their images $A_i$, $B_i$, $C_i$, $D$ which have then to be put to zero to 
ensure closure. The analogue of \lesssimple has now 10 summands, each with
4 ``factors''. Let us consider one of them, the tensor product
$e_3\otimes\beta\otimes\alpha\otimes M_1$. An element of it undergoes now
\eqn\interest{
\eqalign{
e_3\otimes\beta\otimes\alpha\otimes m_1 \,& \mapsto \,
\beta\otimes\alpha\otimes m_1 \,\oplus\,
(- e_3\otimes\beta\alpha\otimes m_1)\,\oplus\,
e_3\otimes\beta\otimes M(\alpha) m_1\cr
&\mapsto\, \big[N(\beta) b_1 - c + b_2 M(\alpha)\big]\,m_1\ .
}}
A more complete analysis reveals that closure amounts to put to zero $a_i$ 
and the quantity inside brackets to zero. This relation reexpresses 
the deformation $c$ of $\beta\alpha$ in terms of deformations $b_i$ of 
$\beta$ and $\alpha$, leaving them unconstrained otherwise. Thus we have
that deformations are given by the two $b_i$, divided by the relations we
already found. This is exactly what we would have done intuitively. 

Actually, for quiver without relations we could have resorted to the more
handy $\Hom$ complex provided by Kac \refs{\kac} (for a review see
\dfrtwo). So the method comes into its own for the case with relations. Let
us then modify the quiver we just saw introducing a relation $\beta\alpha=0$.
We can repeat the computation we just did with a few obvious changes:
in particular the term with a $-$ sign in \interest\ drops, leading instead
to a relation $N(\beta) b_1 + b_2 M(\alpha)= 0$. For $M=N$, this is just
what we would get from the ``naive'' procedure used in the text: considering
\eqn\naive{
(M(\beta) + b_2)(M(\alpha) + b_1)=0
}
to first order (zero order is obviously the representation of the relation).

For other finite quivers (having a finite number of paths) these computations
are not really more difficult. In particular, one can check that for the
Beilinson quiver the standard complex can be simplified to yield the simpler
procedure considered in \dfrtwo. 

The fact that our quivers are not finite, instead, would make
repeating this procedure explicitly in our case would be instead
awfully complicated.  Even considering only the portion we are using
of the McKay quiver (only three consecutive nodes are present), the
relations derived from the superpotential do not make finite the
possible paths: indeed there are for instance loops like $y_{45}
x_1'x_1$, and so on.  So already writing the simplest thing,
\simpleHom, would consist in this case of infinite pieces in each of
the three nodes (indeed because of loops like the one just
mentioned). However, the principles we abstracted from the finite
examples we considered in some detail here are that:

1 - a deformation of a composite path can be reexpressed in terms of 
deformations of the constituents (compare the square bracket in \interest)

2 - when there is a relation, its effect can be reckoned as coming from the 
simple intuitive procedure \naive.

In particular first principle allows us to consider only deformations of 
generators of the possible paths, if we take care of the relations according
to second principle.

\appendix{B}{The topological LG model}

Consider a LG model with chiral superfields $\Phi^i$ and a quasi-homogeneous
superpotential $W(\Phi)$. As shown by Witten, this model can be obtained
as a phase of the gauged linear sigma model. For LG models on worldsheets
with boundary preserving B-type supersymmetry, it has been shown\govjs\hiv\ 
that the
only boundary conditions are those compatible with the condition $W=0$.
In simple models involving a single chiral field, the only possible condition
is the Dirichlet one. This carries over to the case of several chiral
superfields when one imposes boundary conditions separately on each of
the chiral superfields. For LG models which flow in the infrared to
Gepner models associated with Calabi-Yau compactifications, this implies
that all the Recknagel and Schomerus boundary states must necessarily arise from
Dirichlet conditions being imposed on each of the chiral superfields.
Further, when the superpotential is degenerate at $\phi^i=0$, the
condition $W=0$ implies that the RS states arise from the
boundary condition $\phi^i=0$. The boundary conditions that we will consider
in the LG model is given by
\eqn\bc{
\phi^i=0\quad,\quad\tau^i \equiv (\psi^i_+ - \psi^i_-) =0\ , 
}
where the fermionic boundary condition is obtained by  closure
under the unbroken supersymmetry. Let us parametrise the unbroken
supersymmetry $\epsilon=\sqrt2\epsilon_+=\sqrt2\epsilon_-$ under
which the other components of the chiral superfield transform as
\eqn\btrans{\eqalign{
\delta \xi^i &= 2i \bar{\epsilon} \partial_y\phi^i \cr
\delta (\partial_y \phi^i) &= -i\epsilon \partial_x \xi^i\ ,
}}
where $\xi^i\equiv (\psi_+^i + \psi_-^i)$. It follows that
$(\xi^i,\sqrt2 i\partial_x \phi^i)$ form an antichiral fermi
superfield (in boundary superspace with 
$\theta\equiv -\sqrt2 \theta^+ = \sqrt2 \eta \theta^-$)
\eqn\bsuper{
\Xi^i = \xi^i-2\bar{\theta}i \partial_y \phi^i -\theta \bar{\theta}\partial_x
\xi^i\ .
}
Note that we have made use of the fact that $W$ is degenerate at $\phi^i=0$
and thus one has $F_i=0$ on the boundary. Further, we also need to use
the bulk equations of motion for the fermions.

Let us now specialise to the case of the quintic where we have five
chiral superfields $\Phi^1,\cdots,\Phi^5$ and a degree five superpotential.
The boundary interactions take the form
\eqn\bint{
\int d\theta \left( X^i\ \bar{\Xi}^i +
Y^{[ijk]}\ \bar{\Xi}^i\bar{\Xi}^j\bar{\Xi}^k
+ Z\ \bar{\Xi}^1\bar{\Xi}^2\bar{\Xi}^3\bar{\Xi}^4\bar{\Xi}^5 \right)
+h.c.
}
$X$, $Y$ and $Z$ are Chan-Paton matrices.
The assignment of appropriate Chan-Paton factors is not immediately
obvious from the topological LG considerations above. This requires
a careful consideration of allowed fractional brane states
and the states appearing in open-strings connecting them as in \dd.

\subsec{The topological model}

Consider the topological twist where $\bar{\epsilon}$ becomes a scalar.
It follows that in this topological theory, terms which appear in the
action as the following superspace integrals $\int d^4\theta$ or
$\int d^2\bar{\theta}$ or $\int d\bar{\theta}$ are BRST exact. Hence,
the topological theory depends solely on the holomorphic superpotential
$W(\phi)$ in the bulk and ${\cal X}^i$, ${\cal Y}^{ijk}$ and ${\cal Z}$
couplings in the boundary.

In the bulk, any holomorphic function such as $W(\phi)$ is a topological
observable  which is a zero-form operator $O_{(W;0)}$ in the bulk. Its one-form
and two-form versions are obtained by the usual descent procedure.
They are given by
\eqn\bulkobs{\eqalign{
O_{(W;1)} &= {{\partial W}\over{\partial \phi^i}} \tau^i \quad{\rm and}\quad 
{{\partial W}\over{\partial \phi^i}} \xi^i \cr
O_{(W;2)} &= {{\partial^2 W}\over{\partial\phi^i \phi^j}} \tau^i \xi^j
+ 2 \left|{{\partial W}\over{\partial \phi^i}}\right|^2\ ,
}}
where $\tau^i\leftrightarrow \partial_x \phi^i dx$ and 
$\xi^i\leftrightarrow \partial_y \phi^i dy$.

On the boundary, analogously, one has zero-form and one-form observables.
They are
\eqn\boundobs{\eqalign{
O^{[i_1\ldots i_k]}_{(k;0)} &= \bar{\xi}^{i_1}\ldots \bar{\xi}^{i_k} \cr
O^{[i_1\ldots i_k]}_{(k;1)} &= 
2i\partial_y \bar{\phi}^{[i_1}\bar{\xi}^{i_2}\ldots \bar{\xi}^{i_k]}
}}
The index $k$ indicates that they correspond to $k$-forms
$d\bar{\phi}^{i_1}\ldots d\bar{\phi}^{i_k}$ on $\BC^5$. Only odd-forms
survive the GSO projection and these are boundary condition changing
operators (except when $k=5$).

The ghost number in the topological theory is the same as the unbroken
$U(1)_R$ charge. The ghost numbers  assignments are
$$\matrix{
{\rm Field} & \phi & \tau & \xi & F \cr
{\rm Ghost\ No.} & (0+{2\over5}) & (-1+{2\over5}) & (-1+{2\over5}) & (-2+{2\over5}) 
}
$$
where the shift of $2/5$ is the modification due to the addition of a
homogeneous  superpotential of degree five\wittenphases.

\listrefs

\end